\documentclass[twoside,11pt]{article}

\usepackage{blindtext}

\newtheorem{assumption}{Assumption}
\newtheorem{result}{Result}
\usepackage{amsmath}
\usepackage{amsfonts}
\usepackage{psfrag,epsf}
\usepackage{enumerate}
\usepackage{adjustbox}
\usepackage{caption}
\usepackage{subcaption}
\usepackage{multirow}
\usepackage[compact]{titlesec}
\usepackage{comment}
\usepackage[utf8]{inputenc}
\usepackage{listings}
\usepackage{color}

\newcommand{\ig}[1]{{\color{black}{#1}}}
\usepackage{algorithm}
\usepackage{jmlr2e}

\usepackage{lastpage}

\ShortHeadings{Subsampled residual bootstrap}{Ganguly, Sengupta and Ghosh}
\firstpageno{1}

\begin{document}

\title{Scalable Resampling in Massive Generalized Linear Models via Subsampled Residual Bootstrap}

\author{\name Indrila Ganguly \email igangul2@ncsu.edu \\
       \addr Department of Statistics\\
       North Carolina State University\\
       Raleigh, NC 27695-7103, USA
       \AND
       \name Srijan Sengupta
       \email ssengup2@ncsu.edu \\
       \addr Department of Statistics\\
       North Carolina State University\\
       Raleigh, NC 27695-7103, USA
       \AND
       \name Sujit Ghosh \email sujit.ghosh@ncsu.edu \\
       \addr Department of Statistics\\
       North Carolina State University\\
       Raleigh, NC 27695-7103, USA}


\maketitle

\begin{abstract}
 Residual bootstrap is a classical method for statistical inference in regression settings.
 With massive data sets becoming increasingly common, there is a demand for computationally efficient alternatives to residual bootstrap.
 We propose a simple and versatile scalable algorithm called subsampled residual bootstrap (SRB) for generalized linear models (GLMs), a large class of regression models that includes the classical linear regression model as well as other widely used models such as logistic, Poisson and probit regression.
 We prove consistency and distributional results that establish that the SRB has the same theoretical guarantees under the GLM framework as the classical residual bootstrap, while being computationally much faster.
  We demonstrate the empirical performance of SRB via simulation studies and a real data analysis of the Forest Covertype data from the UCI Machine Learning Repository.

\end{abstract}


\section{Introduction}

\label{sec:intro}

As data acquisition technologies advance, we are witnessing an influx of massive data sets across scientific applications. This shift challenges traditional statistical methodologies, as they often prove computationally infeasible for these large-scale data sets.
To quote \cite{jordan2013statistics}, a critical question arising in this massive data era is: ``Can you guarantee a certain level of inferential accuracy
within a certain time budget even as the data grow in size?''
Computational scalability has become a pivotal criterion in modern statistical methodology, to be considered alongside the classical theoretical criteria such as consistency and convergence rates.

Generalized Linear Models (GLMs) have emerged as a cornerstone in supervised statistical learning for massive data sets \citep{wang2018optimal}.
Consider a response variable $Y=(Y_1, \ldots, Y_n)'$ such that each $Y_i$ follows an exponential family distribution with density function given by
\begin{equation}
    f_i(Y_i|\theta_i, \phi_i)= \exp \left(\frac{Y_i\theta_i-b(\theta_i)}{a(\phi_i)}+c(Y_i,\phi_i) \right),
    \label{eq:glm}
\end{equation}
with $E(Y_i)=\mu_i=b'(\theta_i)$ and $Var(Y_i)=a(\phi_i)b''(\theta_i)$.
Let $X=(X_1, \ldots, X_n)'=(X_{.1},\ldots, X_{.p})$ be the $n\times p$ matrix of predictors or features,
and let $\eta_i=X_i'\beta$ be a linear function of the predictors.
Under the GLM framework, the relationship between input variables (predictors or features) and output variables (responses) is modeled by a link function $g$
such that 
$$\eta_i=X_i'\beta=g(\mu_i)$$
for $i=1, \ldots, n$,
and model fitting is usually performed by the
 iteratively reweighted least squares algorithm \citep{green1984iteratively,dobson2018introduction}.
 GLMs are popular due to their versatility in accommodating diverse data distributions and response types, from linear models to logistic regression, Poisson regression, and probit regression \citep{mccullagh2019generalized}.

After estimating the parameters of a GLM,
we are often interested in downstream inference tasks such as hypothesis testing, uncertainty quantification (e.g., via confidence intervals), and quantifying the precision of the estimator.
This typically involves a functional of the sampling distribution of \textit{root function} of the form $T_n(\hat{\beta}, \beta)$.
For example, suppose we want to quantify the precision of the estimator $\hat{\beta}$ by a high quantile (say, 99\%) of the $l_2$ loss.
Then the root function is given by $T_n(\hat{\beta}, \beta) = ||\hat{\beta} - \beta||_2$, and the functional of interest is the $99\%$ quantile of the sampling distribution of the root function.
Similarly, in hypothesis testing, the root function is given by the test statistic under the null, and the functional of interest --- the critical value of the test --- is the $(1-\alpha)$ quantile of the null sampling distribution of the root function where $\alpha$ is the nominal significance level.
Bootstrap resampling \citep{efron1979} is often used to 
estimate the functional of interest by approximating the sampling distribution of $T_n$ by the empirical distribution of the resampled estimate.
Bootstrap stands out for its methodological versatility and automatic nature, as the practitioner simply needs to implement the estimator on resamples of the data (in the same manner as the original estimate), which does not require advanced statistical knowledge unlike using asymptotic distributional results.
Furthermore, bootstrap has excellent theoretical properties such as consistency and higher-order accuracy under quite general settings \citep{singh1981asymptotic,shao2012jackknife,beran1991asympotic, davison1997bootstrap,efron1994introduction,hall1993edgeworth,shao1995jackknife,lahiri2003resampling,chatterjee2011bootstrapping,hall2013simple,senguptadrw,lopes2014residual}. 

Despite these strengths, traditional bootstrap methods are not computationally feasible for massive GLMs,
as each bootstrap iteration involves 
a computational cost of the
same order as that of the original inference on the data \citep{jordan2013statistics, kleiner2014scalable, sengupta2016subsampled}.
Although parallel computing platforms can partially alleviate this problem, they are still computationally very demanding for repeatedly processing massive resampled data sets. 



In this paper, we aim to address this gap 
by developing a 
new procedure called \textit{subsampled residual bootstrap} (SRB).
This method preserves the methodological and theoretical strengths of classical residual bootstrap \citep{freedman1981bootstrapping}, while ensuring computational efficiency for massive GLMs. 
The key idea of SRB is to construct full-size resamples by concatenating smaller subsamples,  instead of directly generating a full-size resample from the empirical distribution of model residual
(see Figure \ref{fig:flowchart} for a schematic diagram).
This simple but powerful modification markedly reduces the computational complexity while retaining the core idea of constructing a resample that acts as a statistically valid proxy of the original sample, and preserving the methodological simplicity of classical residual bootstrap.
Our theoretical analysis establishes that SRB has consistency and asymptotic normality under linear models and GLMs.
The theoretical properties of residual bootstrap follow as a special case of these results.
This is especially significant
for the classical residual bootstrap under the GLM, where, to our knowledge, the theoretical properties were previously unknown.
Thus, from both theoretical and methodological viewpoints, the proposed SRB can be interpreted as a generalization of the classical residual bootstrap.
In this interpretation, our contribution can be viewed as making the bootstrap resampling toolbox for GLMs more flexible by offering a range of options concerning computational scalability, with the existing residual bootstrap being at the slowest end of this range.

The rest of the paper is organized as follows. 
Section \ref{sec:lit} provides a brief review of related work.
In Section \ref{sec:mlr}, we introduce a special case of the SRB procedure under the linear model for clarity and simplicity, 
followed by its full version under the general GLM setting in Section \ref{sec:glm}. 
We describe the theoretical results for SRB under both linear models and GLMs in   Section \ref{sec:theory}. 
We demonstrate the performance of our method under three GLM settings --- linear models, logistic regression, and Poisson regression --- via simulation studies in Section \ref{sec:sim_mlr}.
In Section \ref{sec:real}, we report a case study on the Forest Covertype data from the UCI Machine Learning Repository.
The paper concludes with a discussion in Section \ref{sec:disc}.
All technical proofs are in the Appendix.


\subsection{A brief review of related work}
\label{sec:lit}

There has been a rich development of resampling methods for regression models over the last four decades.
In particular, 
 three popular bootstrapping approaches have been developed for regression settings: paired bootstrap, residual bootstrap \citep{freedman1981bootstrapping}, 
and wild bootstrap \citep{freedman1981bootstrapping, wu1986jackknife, liu1988bootstrap,mammen1993bootstrap}. 
Paired bootstrap applies to the correlation model where the features or predictors are considered random.
For fixed design matrices in regression models, residual bootstrap applies when the errors are homoskedastic and wild bootstrap applies when the errors are heteroskedastic.
All three methods were initially developed under the classical linear regression model.
\cite{moulton1991bootstrapping} extended both paired and residual bootstrap to GLMs, 
but did not provide any theoretical results.
More recently, \cite{chatterjee2011bootstrapping} extended these ideas to high-dimensional regression models, and \cite{eck2018bootstrapping} extended them to multivariate responses.
Our work contributes to this research topic by proposing a scalable alternative to residual bootstrap for massive GLMs.


In related work, recent years have seen a number of notable contributions to the scalable resampling literature.
The classical bootstrapping approach is computationally infeasible for massive data sets since the computational cost for each bootstrap iteration is in the same order as that of the original sample \citep{kleiner2014scalable}.  
Several computationally efficient alternatives to classical bootstrap have been proposed to address this issue, such as $m$ out of $n$ bootstrap  \citep{bickel2012resampling}, subsampling \citep{politis1999subsampling}, and more recently the bag of little bootstraps \citep{kleiner2014scalable}, the subsampled double bootstrap \citep{sengupta2016subsampled}, and
distributed bootstrap \citep{yu2020simultaneous,volgushev2019distributed, chen2021distributed}.
However, none of these methods work for residual bootstrap under generalized linear regression models.
This work fills this crucial gap in the scalable bootstrap toolbox.

\section{Subsampled Residual Bootstrap for Linear Models}
\label{sec:mlr}
Consider the linear model 
\begin{equation}Y = X\beta + \epsilon
\label{eq:regmodel}
\end{equation} 
where $Y=(Y_1, Y_2,\ldots, Y_n)$ is a vector of responses of length $n$, $X=(X_1, \ldots, X_n)'$ is the $n \times p$ design matrix whose $(i,j)^{th}$ element, denoted by $X_{ij}$, is the value of the $j^{th}$ feature for the $i^{th}$ observation,  and $X_i'$ denotes the $i^{th}$ row of $X$, $\beta=( \beta_1, \ldots, \beta_p)'$ is the vector of coefficients and $\epsilon=(\epsilon_1, \ldots, \epsilon_n)'$ is an $n$-dimensional vector of error terms with mean 0 and finite variance $\sigma^2$. 
Here $Y$ and $X$ are observed, while all other terms are unobserved.
Then $\hat{\beta}$, the least squares estimator for $\beta$,  is defined as 
\begin{equation}\hat{\beta} = (X'X)^{-1}X'Y,
\label{eq:regbetahat}
\end{equation} 
and the vector of residuals is given by $\hat{\epsilon}=(\hat{\epsilon}_1, \ldots, \hat{\epsilon}_n)'=Y-X\hat{\beta}$. Here, we have assumed that $n>p$ and that $X$ has full column rank.

\subsection{Residual Bootstrap under the linear model}
We now introduce the classical residual bootstrap (RB) under the linear model.
Note that these $\hat{\epsilon}_i^{'}s$ may not add up to $0$, as the column space of $X$ may not include the constant vector. 
To address this, \cite{freedman1981bootstrapping} suggested resampling from the centered residuals, \ig{$\hat{r}_i=$} $\hat{\epsilon}_i-\frac{1}{n}\sum_{i=1}^n \hat{\epsilon}_i$, before resampling from them. 
Although this idea of resampling from the uncorrected residuals is popular, \cite{el2018can} showed that there remains some discrepancy in the distributions of $\epsilon$ and $\hat{\epsilon}$.
To make the variances of the residuals match those of the true errors, we consider the modified residuals, given by 
                    \begin{equation*}
          \hat{\epsilon}^{(m)}_i=\frac{\hat{\epsilon}_i}{(1-h_i)^{1/2}},
          \end{equation*}
where $h_i$ denotes the $i^{th}$ diagonal element of the projection matrix $X(X'X)^{-1}X'$, and 
compute the centered  modified residuals \citep{davison1997bootstrap}, given by 
\begin{equation*}
\hat{\epsilon}^{(cm)}_i=\hat{\epsilon}^{(m)}_i-\frac{1}{n}\sum_{i=1}^n \hat{\epsilon}^{(m)}_i.
\label{eq:cenmodres}
\end{equation*}
We generate a bootstrap resample $\epsilon^*=(\epsilon_1^*, \ldots, \epsilon_n^*)$ by resampling from the empirical distribution of the centered modified residuals $\{\hat{\epsilon}^{(cm)}_i\}_{i=1}^n$, which is why this procedure is called residual bootstrap. 
Next, the resampled response $Y^*$ is generated as $Y^*=X\hat{\beta}+\epsilon^*$.
The resampled estimate 
is computed as 
\begin{equation}\hat{\beta}^* = (X'X)^{-1}X'Y^*,
\label{eq:regbetastar}
\end{equation} 
and the error variance $\sigma^2$ is estimated as 
\begin{equation}
\hat{\sigma}_n^{*2}=\frac{1}{n} \sum_{i=1}^n \hat{\epsilon}_i^{*2}- \left(\frac{1}{n} \sum_{i=1}^n \hat{\epsilon}_i^*\right)^2,
\end{equation}
where $\hat{\epsilon}^*=(\hat{\epsilon}_1^*,\ldots, \hat{\epsilon}_n^*)'=Y^*-X\hat{\beta}^*$ is the vector of ``starred'' residuals.
We compute $T_n(\hat{\beta}^*, \hat{\beta})$ as a proxy for  $T_n(\hat{\beta}, {\beta})$, with the resampled estimate $\hat{\beta}^*$ acting as a proxy for the sample estimate $\hat{\beta}$, and  the sample estimate $\hat{\beta}$ acting as a proxy for the unknown true value of the coefficient,  ${\beta}$.
This process is repeated $R$ times to obtain $R$ bootstrap \textit{replicates} of $T_n(\hat{\beta}^*, \hat{\beta})$.
We use the empirical distribution from these $R$ replicates to estimate the unknown sampling distribution of the root function $T_n(\hat{\beta}, \beta)$.



\subsection{Subsampled Residual Bootstrap under the linear model}
Computing each resampled estimate $\hat{\beta}^*$ via Equation \eqref{eq:regbetastar} consumes the same resources as computing the original sample estimate $\hat{\beta}$ via Equation \eqref{eq:regbetahat}. 
Residual bootstrap requires $R$ computations of the same order, where $R$ needs to be large enough such that the empirical distribution is a good estimate of the true distribution of $T_n(\hat{\beta}^*, \hat{\beta})$ conditional on $X$ and $Y$.
This becomes computationally infeasible for massive data sets. 


Consider a simple but useful paraphrasing of Equation \eqref{eq:regbetastar}.
\begin{equation}\hat{\beta}^* = (X'X)^{-1}X'Y^*
= (X'X)^{-1}X'(X\hat{\beta}+\epsilon^*)
=\hat{\beta}+(X'X)^{-1}X'{\epsilon}^*
\label{eq:regbetastar2}
\end{equation} 
The advantage of the final expression is that $\hat{\beta}$ and $(X'X)^{-1}X'$ need to be computed just once as overhead and stored in memory.
Then, to compute $\hat{\beta}^*$ for each RB iteration, we only need to multiply the stored matrix $(X'X)^{-1}X'$ with the resampled residuals ${\epsilon}^*$ for that specific iteration, and add the product $(X'X)^{-1}X'{\epsilon}^*$ to the already stored $\hat{\beta}$.

This interpretation does not offer any computational benefit to the residual bootstrap method itself since multiplying $(X'X)^{-1}X'$  with ${\epsilon}^*$ has the same computational complexity as computing $(X'X)^{-1}X'Y^*$ following Equation \eqref{eq:regbetastar}.
But this interpretation is pivotal to the proposed SRB method.
 Under the SRB, instead of generating a resample of size $n$,
 we generate a small subsample of size $b<n$ (usually $b=o(n)$), denoted by $\epsilon_{(b)}^{*}=(\epsilon_{1}^{*}, \ldots, \epsilon_{b}^{*} )$. Let $m=n/b$ be a natural number for notational convenience. Further, let $\epsilon_{(b)}^{**}$ be a vector of length $n$ formed by concatenating $\epsilon_{(b)}^{*}$ $m$ times, i.e.,
 $\epsilon_{(b)}^{**} = (\epsilon_{1}^{*}, \ldots, \epsilon_{b}^{*}, \ldots, \epsilon_{1}^{*}, \ldots, \epsilon_{b}^{*} )$. 
 The key idea of SRB is to use $\epsilon_{(b)}^{**}$, this concatenated n-length vector, as a``full-size resample'' instead of resample of length $n$ directly sampled from $\{\hat{\epsilon}^{(cm)}_i\}_{i=1}^n$. 

 How does this help?
 From a statistical perspective, $\epsilon_{(b)}^{**}$ behaves like a random sample from the empirical distribution of centered and modified residuals.
 As before, $Y^*$ is generated as $Y^*=X\hat{\beta}+\epsilon_{(b)}^{**}$.
 Note that $\epsilon_{(b)}^{**}=J'\epsilon_{(b)}^{*}$ where $J_{b \times mb} = (\mathbb{I}_{b \times b} \cdots \mathbb{I}_{b \times b})$ is the ${b \times mb}$ matrix formed by concatenating $m$ identity matrices of order $b$ row wise. 
  Then, mimicking Equations \eqref{eq:regbetastar} and \eqref{eq:regbetastar2}, the SRB estimator is given by  \begin{equation}
  \hat{\beta}_{(b)}^{*}= 
  (X'X)^{-1}X'Y^*
  =
  (X'X)^{-1}X'(X\hat{\beta}+\epsilon_{(b)}^{**})
  =
  \hat{\beta}+(X'X)^{-1}X'J'\epsilon_{(b)}^{*}.
 \label{eq:regbetastarsrb}
 \end{equation}
Thus, from a methodological perspective, SRB works very similarly to RB.
\ig{Note that, analogous to RB, we can compute the residuals based on SRB as: $\hat{\epsilon}^* =(\hat{\epsilon}_1^*, \ldots, \hat{\epsilon}_n^*)= Y^*-X\hat{\beta}_{(b)}^*$. Hence, we can estimate $\sigma^2$ as \begin{equation}
\hat{\sigma}_n^{*2}=\frac{1}{n} \sum_{i=1}^n \hat{\epsilon}_i^{*2}- \left(\frac{1}{n} \sum_{i=1}^n \hat{\epsilon}_i^* \right)^2.
\end{equation}
Nevertheless, the true variance $\sigma^2$ can also be estimated using the residuals obtained after initially fitting the linear model, which is given by \begin{equation}\hat{\sigma}_n^2=\frac{1}{n} \sum_{i=1}^n \hat{r}_i^2- \left(\frac{1}{n} \sum_{i=1}^n \hat{r}_i \right)^2.\end{equation} }

From a computational perspective, 
the computational complexity for each RB iteration is $O(np)$, which consists of multiplying the stored matrix $(X'X)^{-1}X'$ with the resampled residuals ${\epsilon}^*$ for that specific iteration, and adding the product $(X'X)^{-1}X'{\epsilon}^*$ to the already stored $\hat{\beta}$. 
Hence, if $R$ resamples are generated, the computational complexity for RB is $O(npR)$. 
On the other hand, from the final expression in Equation \eqref{eq:regbetastarsrb}, SRB requires the following computations.
First, we need to multiply $(X'X)^{-1}X'$ with $J$ only once and store the product, $(X'X)^{-1}X'J'$, in memory for future use.
Then, for each SRB iteration, we need to compute the product of the stored matrix $(X'X)^{-1}X'J'$ with the resampled residuals $\epsilon_{(b)}^{*}$ for that specific iteration, and add the product $(X'X)^{-1}X'J'\epsilon_{(b)}^{*}$ to the already stored $\hat{\beta}$. 
The computational complexity is therefore $O(bpR)$ for $R$ resamples. 
Thus, the computational complexity for the main resampling step is $O(bpR)$ under SRB compared to $O(npR)$ under RB,
which makes SRB approximately $n/b=m$ times faster. 
This leads to substantial computational savings, especially when $b$ is much smaller than $n$.
For example, suppose $n=10^6$ and $b=n^{2/3}$. Then, SRB is expected to be about 100 times faster than RB for the same number of resamples. 
The key idea is reducing the complexity per iteration from $O(np)$ to $O(bp)$ by using a concatenated subsample instead of a full-size resample. 

    

\noindent\textbf{Remark:} At this juncture, a natural question to ask is whether we can simply use subsampling directly.
Recall that $\epsilon_{(b)}^{*}$ is a subsample of size $b$ from the empirical distribution of the centered and modified residuals.
Instead of constructing the full-size resample $\epsilon_{(b)}^{**}$ by repeatedly concatenating this subsample, why not simply use $\epsilon_{(b)}^{*}$ directly to construct a subsample of the data, of size $b$?
The reason for not doing this is that the subsample behaves like a data set of size $b$ rather than $n$.
Therefore, the root function computed from the subsample will behave as $T_b$ rather than as $T_n$, and the practitioner will need to rescale the output in order to estimate the statistical function of interest.
This reduces the practical convenience of the method, as the practitioner will need to know the convergence rates.
The same issue was identified by \cite{kleiner2014scalable} and \cite{sengupta2016subsampled} in the context of nonparametric bootstrap, and we quote the relevant part from \cite{kleiner2014scalable} here: ``because the variability of an estimator on a subsample differs from its variability on the full
data set, these procedures must perform a rescaling of their output, and this rescaling requires
knowledge and explicit use of the rate of convergence of the estimator in question; these methods
are thus less automatic and easily deployable than the bootstrap.''

\ig{We note that in the special case where the statistic of interest is asymptotically pivotal, subsampling does not require any scaling, and therefore can be applied without the issues described above.
We illustrate the computational efficiency of subsampling in the pivotal case via a simulation study in the Appendix.}

\section{Subsampled Residual Bootstrap for Generalized Linear Models}
\label{sec:glm}
We start with a  description of the GLM estimation framework.
From equation \eqref{eq:glm},
assuming $\phi$ to be constant, 
the log likelihood function expressed as a function of $\mu_i$'s is given by $\sum_{i=1}^n l(\mu_i;Y_i)=\sum_{i=1}^n log f_{i}(Y_i|\theta_i, \phi)$.
The maximum likelihood estimate $\hat{\beta}_j$ for $j=1, \ldots p$ is obtained by solving the likelihood equation \begin{equation}
\sum_{i=1}^n \frac{\partial}{\partial \beta_j} l(\mu_i;Y_i)=0.
\end{equation} 
The likelihood equations can be expressed as \begin{equation}
X'\Delta (Y-\mu)=0
\end{equation} where $\Delta$ is a $n \times n$ diagonal matrix with $\Delta_{ii}=\frac{d \theta_i}{d \eta_i}$. Since $\mu$ is not a linear function of $\beta$, we cannot obtain a closed-form expression for $\hat{\beta}$ like the linear models case. However, the problem can be reformulated as a weighted least squares problem \citep{mccullagh2019generalized}.
Suppose, we define $z_i$ as $(z_i-\eta_i)=(Y_i-\mu_i)\frac{d\eta_i}{d\mu_i} \ \implies z_i=\eta_i+ (Y_i-\mu_i)\frac{d\eta_i}{d\mu_i}$ and perform a least squares regression $z \sim X_{.1}, \ldots, X_{.p}$. Let $V$ be a $n \times n$ diagonal matrix with $V_{ii}=\frac{d\mu_i}{d \theta_i}=Var(Y_i)$. Then $\frac{d \mu}{d \eta}=V \Delta$, and $z= \eta + (V \Delta)^{-1} (Y-\mu)$.
The transformed regression problem is \begin{equation}X'\Delta V \Delta (z-X \beta)=0.\end{equation} We solve this using iteratively re-weighted least squares estimation where the $(t+1)^{th}$ update is given by: 
\begin{equation}\hat{\beta}^{(t+1)}=(X'\hat{\Delta}^{(t)} \hat{V}^{(t)} \hat{\Delta}^{(t)} X)^{-1} X'\hat{\Delta}^{(t)} \hat{V}^{(t)} \hat{\Delta}^{(t)} z^{(t)}.
\end{equation} 
Suppose $\hat{\beta}^{(t)}$ is the current estimate, then we compute $\hat{\eta}^{(t)}=X \hat{\beta}^{(t)}$, $\hat{\mu}^{(t)}=g^{-1}(\hat{\eta}^{(t)})$, $\hat{V}^{(t)}$, $\hat{\Delta}^{(t)}$ at the current estimate and hence compute $z^{(t)}$. We then regress $z^{(t)}\sim X_{.1}, \ldots X_{.p}$ to obtain $\hat{\beta}^{(t+1)}$ and keep iterating until convergence. We denote the final estimate, obtained after convergence, by $\hat{\beta}$.

\subsection{Residual Bootstrap under GLMs}
\cite{lee1990bootstrapping} extended paired bootstrap to the logistic regression setting (which is a special case of GLMs) under the correlation model.
Using an idea similar to residual bootstrap for linear regression models, \cite{moulton1991bootstrapping} proposed the use of  one-step resampling strategies for GLMss under the assumption of deterministic (non-random) regressors. 
  \cite{moulton1991bootstrapping} proposed considering the standardized Pearson residuals $\hat{r}_i'=\frac{\hat{\epsilon}_i}{\sqrt{\hat{v}_i(1-h_i)}}$ where $\hat{v}_i$ is the $i^{th}$ diagonal of $\hat{V}$, $h_i$ is the $i^{th}$ diagonal of the projection matrix $\hat{B}(\hat{B}'\hat{B})^{-1}\hat{B}'$ with $\hat{B}=\hat{V}^{1/2}\hat{\Delta} X$ (with $\hat{V}$ and $\hat{\Delta}$ estimated at the step where the MLE $\hat{\beta}$ is computed) and $\hat{\epsilon}_i=y_i-\hat{y}_i$. Let $\hat{F}_n$ denote the empirical distribution of the standardized Pearson residuals. Then, we generate a bootstrap sample of size $n$ $\epsilon^*= (\epsilon_1^*, \ldots, \epsilon_n^*)$ by resampling from $\hat{F}_n$. Note that, although resamples are generated by drawing from iid samples in Bootstrap methods, such residuals are not available for GLM. Hence , the nearly exchangeable standardized Pearson residuals are used. 

After obtaining the vector of resamples, analogous to linear models, we obtain \begin{equation}\hat{\beta}^{*} = \hat{\beta}+(\hat{B}'\hat{B})^{-1}\hat{B}'{\epsilon}^{*}.\end{equation} 
The variance of the estimated bootstrap coefficient is obtained as follows: \begin{equation}\text{Var}(\hat{\beta}^*)=\left[\frac{1}{n}\sum_{i=1}^n \hat{r}_i^2-\frac{1}{n^2}(\sum_{o=1}^n \hat{r}_i)^2 \right](\hat{B}'\hat{B})^{-1}\end{equation} Hence, the residual bootstrap estimate of $\text{Var}( \hat{\beta})$ can be obtained based on the available data.

\subsection{Subsampled Residual Bootstrap under GLMs}

\begin{figure}[h]
    \centering
    \includegraphics[width=0.9\textwidth]{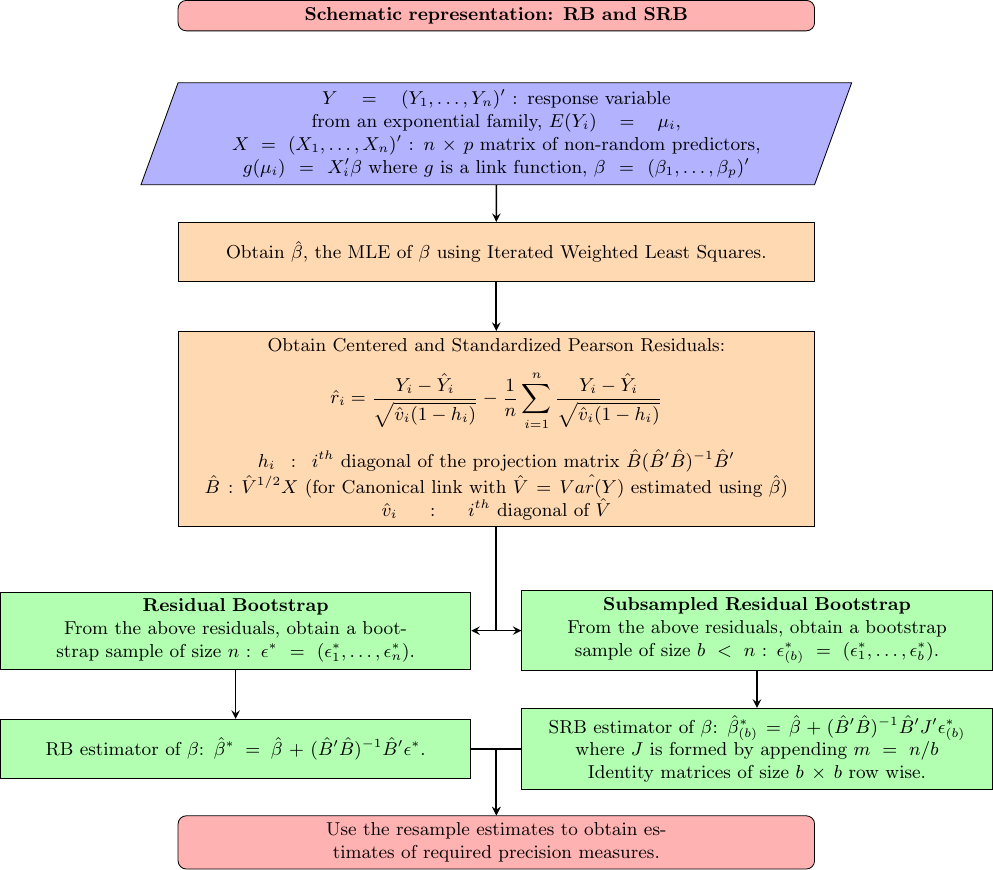}
    \caption{Comparison of Residual Bootstrap and Subsampled Residual Bootstrap methods for GLMs}
    \label{fig:flowchart}
\end{figure}

The fundamental idea is the same as that for linear models.
Instead of directly generating a full-size resample of the model residuals, we construct a full-size resample by concatenting smaller subsamples from the model residuals.
We consider the centered standardized Pearson residuals given by: \begin{equation}\hat{r}_i =\frac{\hat{\epsilon}_i}{\sqrt{\hat{v}_i(1-h_i)}}- \frac{1}{n} \sum_{i=1}^n \frac{\hat{\epsilon}_i}{\sqrt{\hat{v}_i(1-h_i)}}\end{equation} for $i=1,\ldots, n$. Let $\hat{H}_n$ denote the empirical distribution of the centered standardized Pearson residuals. Then, we generate a bootstrap sample of size $b<n$, denoted by $\epsilon_{(b)}^*= (\epsilon_1^*, \ldots, \epsilon_b^*)$ by resampling from $\hat{H}_n$. Hence, a subsampled version of the residual resampling scheme of \cite{moulton1991bootstrapping} can be obtained analogously for GLMs, and we can compute \begin{equation}
    \hat{\beta}_{(b)}^{*}=\hat{\beta}+(\hat{B}'\hat{B})^{-1}\hat{B}'J'{\epsilon}_{(b)}^{*}\end{equation} 
    with the notations described earlier. For GLMs with natural link functions, the equation can be written as: \begin{align*}
    \hat{\beta}_{(b)}^*&=\hat{\beta}+ (X'\hat{V}X)^{-1} X'\hat{V}^{1/2}J'\epsilon_{(b)}^{*} \\ &= \hat{\beta}+ (X'\hat{V}X)^{-1} \Theta'\epsilon_{(b)}^{*}  \\ &= \hat{\beta}+ (F_n(\hat{\beta}))^{-1} \Theta' \epsilon_{(b)}^{*} \end{align*} where $\Theta'=X'\hat{V}^{1/2}J'$ and $F_n(\hat{\beta})=X'\hat{V}X$.
    Figure \ref{fig:flowchart} provides a schematic diagram of SRB and RB.
Note that SRB for the linear model is a special case of SRB under the more general GLM setting.

From a computational perspective, the benefits of SRB are similar to the linear model setting.
Note that $(\hat{B}'\hat{B})^{-1}\hat{B}'$ needs to be computed once and is common to both RB and SRB. Once it is computed, the computational complexity of $(\hat{B}'\hat{B})^{-1}\hat{B}'\epsilon^*$ is $O(np)$ for each SRB resample. Hence, if $R$ samples are considered, the computational complexity is $O(npR)$. On the other hand, for SRB, the computational complexity of $(\hat{B}'\hat{B})^{-1}\hat{B}'J'$ is $O(np)$ and once it is computed, the computational complexity of $(\hat{B}'\hat{B})^{-1}\hat{B}'J'\epsilon_{(b)}^*$ is $O(bpR)$ for $R$ resamples. Thus, computational time reduces considerably, more specifically, becomes approximately $n/b=m$ times faster. This is significant, especially in the scenario when $n>>b$.


\section{Asymptotic Theory for Subsampled Residual Bootstrap}
\label{sec:theory}

In this section, we investigate the theoretical properties of subsampled residual Bootstrap for both linear models and GLMs.
Note that consistency of the classical residual bootstrap under linear models was established by 
\cite{freedman1981bootstrapping}.

\subsection{Linear Models}
\label{subsec:theory_mlr}
We consider the set-up introduced in Section \ref{sec:mlr}, and state the required assumptions.
We will use the notation $a_n >> b_n$ to denote that $a_n/b_n \rightarrow \infty$ as $n \rightarrow \infty$.
\begin{assumption} \label{assum1}
         The matrix $X$ is a fixed design matrix, that is, the elements are non-random. \end{assumption}
         \begin{assumption} \label{assum2}The components $\epsilon_i$ of $\epsilon$, for $i=1,\ldots, n$, are independent and identically distributed with common distribution $F$ with mean 0 and finite unknown variance $\sigma^2$.\end{assumption}

  \begin{assumption} \label{assum_mlr3}
 As $n \rightarrow \infty$, $b \rightarrow \infty$.
 \end{assumption} 
 \begin{assumption} \label{assum_mlr4}
 Let $X'J'JX$ be a $p \times p$ matrix such that $\frac{X'J'JX}{n} \rightarrow Q_b$ where $Q_b$ is a positive definite matrix.
 \end{assumption} 
  \begin{assumption} \label{assum_mlr5} The elements of $X'J'$ are $o(\sqrt{n})$.
 \end{assumption} 

 
\ig{Assumptions 1 and 2 are the same as (1.2) and (1.3) in \cite{freedman1981bootstrapping} and constitute the basic setup under linear models.
Assumptions 4 and 5 are generalizations of the assumptions from \cite{freedman1981bootstrapping}.
We can interpret the classical RB as a special case of SRB where $b=n$ and $J=\mathbb{I}$. 
Then,  (1.4) of \cite{freedman1981bootstrapping} is the corresponding special case of Assumptions 4 and 5, and $X'X$ is a $p \times p$ matrix such that $\frac{X'X}{n} \rightarrow Q_n$ where $Q_n$ is positive definite.}
\ig{We note that choosing $b >> \sqrt{n}$ is sufficient to satisfy assumptions \ref{assum_mlr3}, \ref{assum_mlr4}, and \ref{assum_mlr5}. This holds because the elements of $X$ are all $O(1)$, which imples that a typical element of $X'J'$ is $O(m)$. Recall that $m=n/b$. Thus, Assumption \ref{assum_mlr5} is satisfied for $m=o(\sqrt{n})$, and we have $\sqrt{n}= o(b)$ when $b >> \sqrt{n}$.}

 \begin{theorem} \label{th1} Consider the linear model and suppose that Assumptions \ref{assum1}-\ref{assum_mlr5} hold. Then, conditional on $Y_1, \ldots, Y_n$, $\hat{\beta}_{(b)}^*$  converges in probability to  $\hat{\beta}$ as  $n \rightarrow \infty$. \end{theorem}

 Refer to the Appendix for a proof of Theorem \ref{th1}. This result establishes the consistency of the SRB estimator for Multiple Linear Regression and states that for a particular sample, the SRB estimator of $\beta$ approaches the usual \ig{OLS (ordinary least squares) estimator} of $\beta$ calculated based on the sample, in probability.

 \begin{theorem} \label{th2} Consider the linear model and suppose that Assumptions \ref{assum1}-\ref{assum_mlr5} hold. Then, conditional on \ig{almost surely all sequences} $Y_1, \ldots, Y_n$, \begin{enumerate}
    \item The conditional distribution of $\sqrt{n}(\hat{\beta}_{(b)}^*-\hat{\beta})$ converges weakly to normal with mean 0 and variance-covariance matrix $\sigma^2 Q_n^{-1}Q_bQ_n^{-1}$.
     \item The conditional distribution of the pivot ${(X'J'JX)^{-1/2}(X'X)} (\hat{\beta}_{(b)}^*-\hat{\beta})/\hat{\sigma}_n$ converges to standard normal in $\mathbb{R}^p$. 
   

     \item The conditional distribution of $\hat{\sigma}_n^*$ converges to point mass at $\sigma$.
    \item The conditional distribution of the pivot ${(X'J'JX)^{-1/2}(X'X)} (\hat{\beta}_{(b)}^*-\hat{\beta})/\hat{\sigma}_n^*$ converges to standard normal in $\mathbb{R}^p$.
\end{enumerate}

\end{theorem}
 Refer to the Appendix for a proof of Theorem \ref{th2}. 
 \ig{This result establishes the distributional convergence of the SRB estimator with $\sqrt{n}$-scaling, analogous to well-known classical results for the RB estimator as established in Theorem 2.2 of \cite{freedman1981bootstrapping}.
 Note that the aforementioned classical results for the RB are now subsumed into the above theorem as a special case, by using $b=n$ and $J=\mathbb{I}$.}
 Furthermore, this theorem illustrates that conditional on a given sample, the distribution of ${(X'J'JX)^{-1/2} (X'X)}(\hat{\beta}_{(b)}^*-\hat{\beta})/\hat{\sigma}_n$ or ${(X'J'JX)^{-1/2} (X'X)} (\hat{\beta}_{(b)}^*-\hat{\beta})/\hat{\sigma}_n^*$, computed based on the data, provides a good approximation to that of $(X'X)^{1/2}(\hat{\beta}-\beta)$. Besides, we are able to use the estimate of $\sigma$ computed based on the starred residuals for arriving at similar conclusions as when the estimate based on the original sample is used.


\subsection{Generalized Linear Models}
\label{subsec:theory_glm}

We now look at the consistency results for Subsampled Residual Bootstrap under Generalized Linear Models. For that, we consider the setup introduced in Section \ref{sec:glm}.
We restrict our attention to GLMs with canonical link function. 
Note that the consistency of classical RB under this framework was not known previously, and it follows from our results as a special case of SRB.

We first note that under some standard regularity conditions \citep{fahrmeir1985consistency} and Assumption \ref{assum1}, it can be shown that the unique solution of the likelihood equation $\hat{\beta}$ satisfies \begin{enumerate}
    \item $\hat{\beta} \xrightarrow{a.s} \beta$ as $n \rightarrow \infty$.
    \item $(X' \hat{V} X)^{1/2}(\hat{\beta}-\beta)\xrightarrow{d}N(0, I)$ as $n \rightarrow \infty$.
\end{enumerate}
Furthermore, it can be easily shown using Slutsky's theorem that under Assumption \ref{assum_glm1}, $\sqrt{n}(\hat{\beta}-\beta) \xrightarrow{d} N(0, M_n^{-1})$ where $M_n$ is a positive definite matrix. 
We now start the theoretical analysis of the SRB by stating the following assumptions. 

\begin{assumption} \label{assum_glm1} $\frac{1}{n}X'VX \rightarrow M_n$ which is positive definite.\end{assumption}

\begin{assumption} \label{assum_glm2} $\frac{1}{n}X'V^{1/2} J'J V^{1/2} X \rightarrow M_b$ which is positive definite.\end{assumption} 

\begin{assumption} \label{assum_glm3}The elements of the matrix $X'V^{1/2} J'$ are all $o(\sqrt{n})$.\end{assumption} 

Assumptions \ref{assum_glm2} and \ref{assum_glm3} are specific to Subsampled Residual Bootstrap. Note that when $b=n$, Assumption \ref{assum_glm2} reduces to Assumption \ref{assum_glm1}. Also, note that Assumption \ref{assum_glm3} is a more general version of Assumption \ref{assum_mlr5} used in the results for multiple linear regression where we take $V=\sigma^2 I$.
\ig{Similar to the linear models case, here also  $b>>\sqrt{n}$ is a sufficient condition to ensure that the Assumptions \ref{assum_glm2} and \ref{assum_glm3} are satisfied. Assuming that the elements of $X$ and $V$ are all $O(1)$, we can show that a typical element of $X'V^{1/2}J'$ is $O(m)$. Thus, Assumption \ref{assum_glm3} is satisfied for $m=o(\sqrt{n})$, and since $m=n/b$, we can say $\sqrt{n}= o(b)$.}
We now note two consistency results for SRB in GLMs.



\begin{theorem} \label{th3}  We assume the generalized linear model with Assumptions \ref{assum1},\ref{assum_mlr3},\ref{assum_glm1}-\ref{assum_glm3}, along with the regularity conditions for GLMs and other conditions as specified in \cite{fahrmeir1985consistency}.Then, conditional on $Y_1, \ldots, Y_n$, $\hat{\beta}_{(b)}^*$  converges in probability to  $\hat{\beta}$ as  $n \rightarrow \infty$. \end{theorem}

This result establishes the consistency of the SRB estimator for Generalized Linear Models, stating that conditional on a given sample, the SRB estimator of $\beta$ approaches the usual MLE of $\beta$ calculated based on the sample, in probability. The proof of this theorem is straightforward and follows from the sufficient conditions for convergence in probability. For more details on the proof of \ref{th3}, refer to Appendix. 

\begin{theorem} \label{th4}
 We assume the Generalized Linear model with Assumptions \ref{assum1},\ref{assum_mlr3},\ref{assum_glm1}-\ref{assum_glm3}, along with the regularity conditions for GLMs and other conditions as specified in \cite{fahrmeir1985consistency}. Then, conditional on sample paths $Y_1, \ldots, Y_n$ (with the probability of such sample paths equal to 1), as $n \rightarrow \infty$, \begin{enumerate}
    \item The conditional distribution of $\sqrt{n}(\hat{\beta}_{(b)}^*-\hat{\beta})$ converges weakly to normal with mean 0 and variance-covariance $M_n^{-1}M_bM_n^{-1}$.
    \item The conditional distribution of ${(X'\hat{V}^{1/2}J'J\hat{V}^{1/2}X)^{-1/2}(X'\hat{V}X)}(\hat{\beta}_{(b)}^*-\hat{\beta})$ converges to standard normal in $\mathbb{R}^p$.
\end{enumerate}
\end{theorem}

The proof of this theorem follows by applications of the Lindeberg-Feller Central Limit Theorem and Kolmogorov's SLLN. Refer to the Appendix for a more formal proof of Theorem \ref{th4}. This result establishes the $\sqrt{n}$-consistency of the SRB estimator for GLMs. Note that since the special case of $b=n$ in SRB corresponds to residual bootstrap, this result also establishes $\sqrt{n}$ consistency for the residual bootstrap estimator of regression coefficients in Generalized Linear models, results specific to which could not be found in the literature. Thus, both SRB and RB were found to achieve the same convergence rate. Further, this theorem illustrates that when a particular sample is kept fixed, the distribution of ${(X'\hat{V}^{1/2}J'J\hat{V}^{1/2}X)^{-1/2}(X'\hat{V}X)}(\hat{\beta}_{(b)}^*-\hat{\beta})$, computed based on the data, provides a good approximation to that of $(X'\hat{V}X)^{1/2}(\hat{\beta}-\beta)$.

\section{Simulation Study}
\label{subsec:sim1}
\label{sec:sim_mlr}
We report results from several numerical studies to illustrate the performance of SRB under three GLM settings: Linear models, Logistic regression, and Poisson regression.
We compared SRB with classical RB with respect to computational runtime (measured in seconds) vs. statistical error under each setting.
The root function of interest is $T_n(\hat{\beta},\beta)=\lVert \hat{\beta}-\beta \rVert_2$, and the statistical functional of interest is a quantile, $q$, of the sampling distribution of $T_n$.
The error rate is defined as $\xi = \lvert \hat{q}/q-1 \rvert$ following \cite{kleiner2014scalable} and \cite{sengupta2016subsampled}.
In order to compute $\xi$ for the simulation studies, we numerically approximated $q$ by using a large number of Monte Carlo iterations for each model (5000 for the linear model and 3000 for logistic and Poisson regression).
For each GLM setting, we generated $M=48$ data sets and carried out $B=25$ iterations of SRB and RB  for each data set. This choice of $M$ and $B$ ensures that the standard error of the average error rate is below 0.01 (see the Appendix for a proof).
Each iteration of SRB or RB involves $R=100$ resamples. 
For SRB, we take $b=n^{\gamma}$ with $\gamma \in \{0.5,0.6,0.7,0.8,0.9 \}$. 
\ig{To perform logistic and Poisson regression, we employed the \texttt{glm()} function from the \texttt{stats} package in R, using the default starting values for the iteratively re-weighted least squares procedure.}

\subsection{Linear model}


    Consider a $p$-dimensional multiple regression model $$Y_i = \sum_{j=1}^p \beta_j X_{ij} + \epsilon_i$$ where the parameter of interest is the vector of slope coefficients $\beta=(1, \ldots, 1, 0, \ldots, 0)'$ with half of the elements being equal to 1, and the rest equal to 0. 
    Our
    target precision parameter is the $99 \%$ quantile, $q_{0.99}$, of the true distribution of $T_n(\hat{\beta},\beta) = \lVert \hat{\beta}-\beta \rVert_2$.
    We fixed \ig{$(n,p)= (10^5, 200)$ and $(2\times 10^5, 300)$}, and generated
    $X_{ij}\stackrel{\text{iid}}{\sim} Pareto(\alpha=3)$, $\epsilon_i \stackrel{\text{iid}}{\sim} \chi_1^2-1$. 

\subsection{Logistic regression}
\label{sec:sim_glm_log}

     Consider a $p$-dimensional multiple logistic regression model $$Y_i \stackrel{ind}{\sim} Ber(p_i) \ \text{where} \ \ logit(p_i)=\sum_{j=1}^p \beta_j X_{ij}$$
    Our parameter of interest is the vector of slope coefficients $\beta=(0.2,0.2,0, \ldots, 0)'$. 
  The target precision parameter is the $95 \%$ quantile, $q_{0.95}$, of the true distribution of $T_n(\hat{\beta},\beta)=\lVert \hat{\beta}-\beta \rVert_2$. 
     We fixed \ig{$(n,p)= (50000, 200)$ and $(10^5, 300)$}, 
    and generated $X_{ij}\stackrel{\text{iid}}{\sim} U(-1,1)$.

\subsection{Poisson regression}
\label{sec:sim_glm_poi}
Finally, consider a $p$-dimensional poisson regression model $$Y_i \stackrel{ind}{\sim} Poisson(\lambda_i) \ \text{where} \ \ log(\lambda_i)=\sum_{j=1}^p \beta_j X_{ij}$$ with the vector of slope coefficients $\beta=(0.2,0.2,0, \ldots, 0)'$.
We fixed \ig{$(n,p)= (50000, 200)$ and $(10^5, 300)$}, 
    and generated $X_{ij}\stackrel{\text{iid}}{\sim} U(-1,1)$.
    The target precision parameter is the $95 \%$ quantile, $q_{0.95}$, of the true distribution of $T_n(\hat{\beta},\beta)=\lVert \hat{\beta}-\beta \rVert_2$. 


\subsection{Results}
The average error rates (with standard deviations) and runtimes are tabulated in \ig{Tables \ref{tab:sim_results} and \ref{tab:sim_results_big}}, and the detailed individual outcomes are plotted in \ig{Figures \ref{fig:sim} and \ref{fig:sim_big}}.
We observe that SRB with $\gamma > 0.5$ is computationally much more efficient than RB.
The proposed method provides estimates of the target statistical functional, $q$, that are statistically as accurate (and in some cases even more accurate) as RB within runtimes that are orders of magnitude smaller than that for RB.
The benefits of SRB over RB hold persistently across the three GLM settings: linear models, logistic regression, and Poisson regression, which
establishes that SRB can be used as a scalable alternative to RB across the range of GLMs without much loss of statistical accuracy.

Although the theoretical results require $b >> n^{0.5}$ for accuracy guarantees, we included 
the $b = n^{0.5}$ case to study it from a numerical perspective.
We observe that the errors for $b = n^{0.5}$ are indeed much higher than $b >> n^{0.5}$ for linear models and Poisson regression, which shows that our theoretical results provide useful guidelines for numerical performance.

\ig{
\begin{figure}[h]
    \centering
    \includegraphics[trim={0cm 3cm 0cm 0cm},clip, scale=0.38]{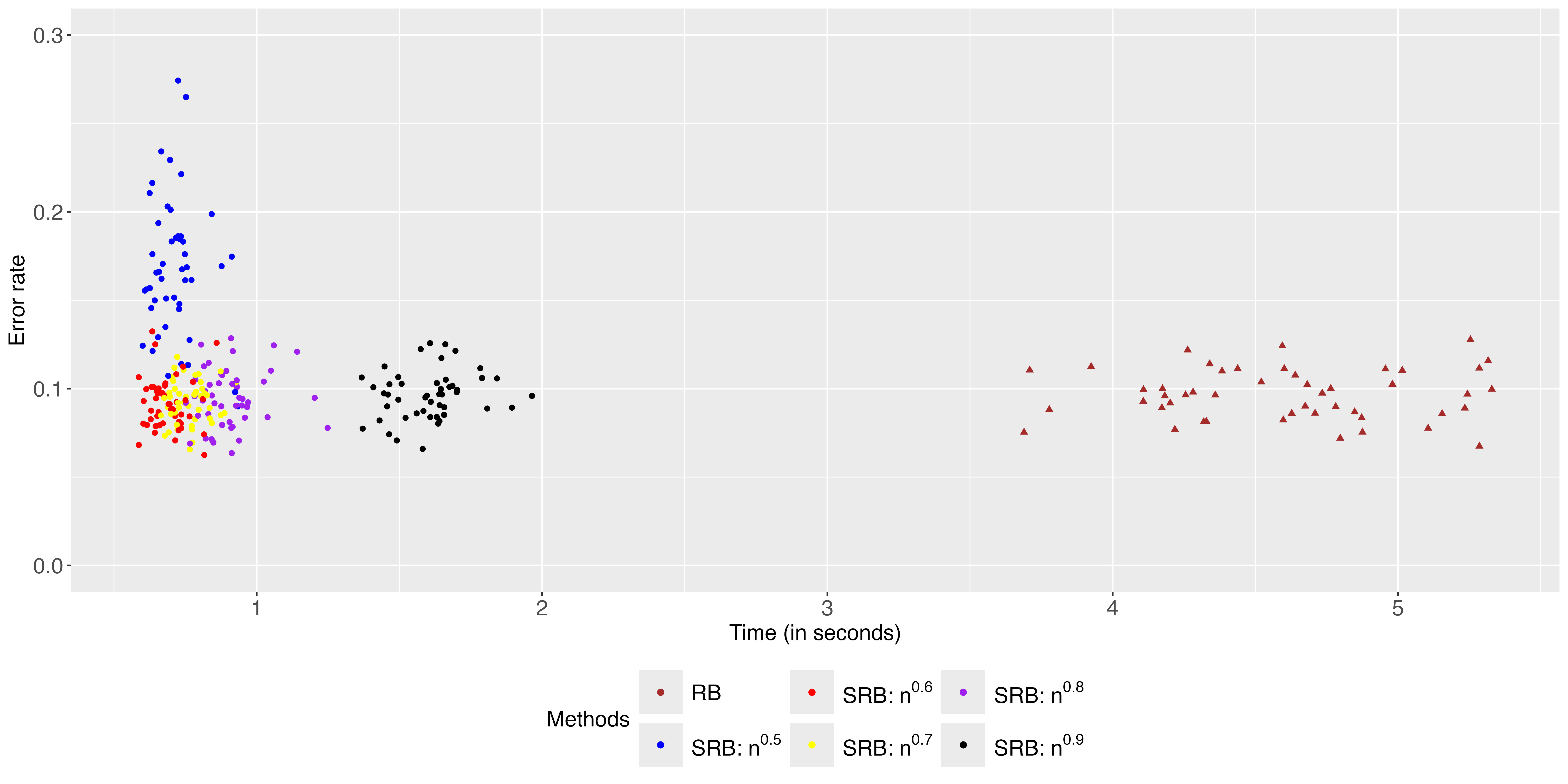}\\
    \includegraphics[trim={0cm 3cm 0cm 0cm},clip, scale=0.38]{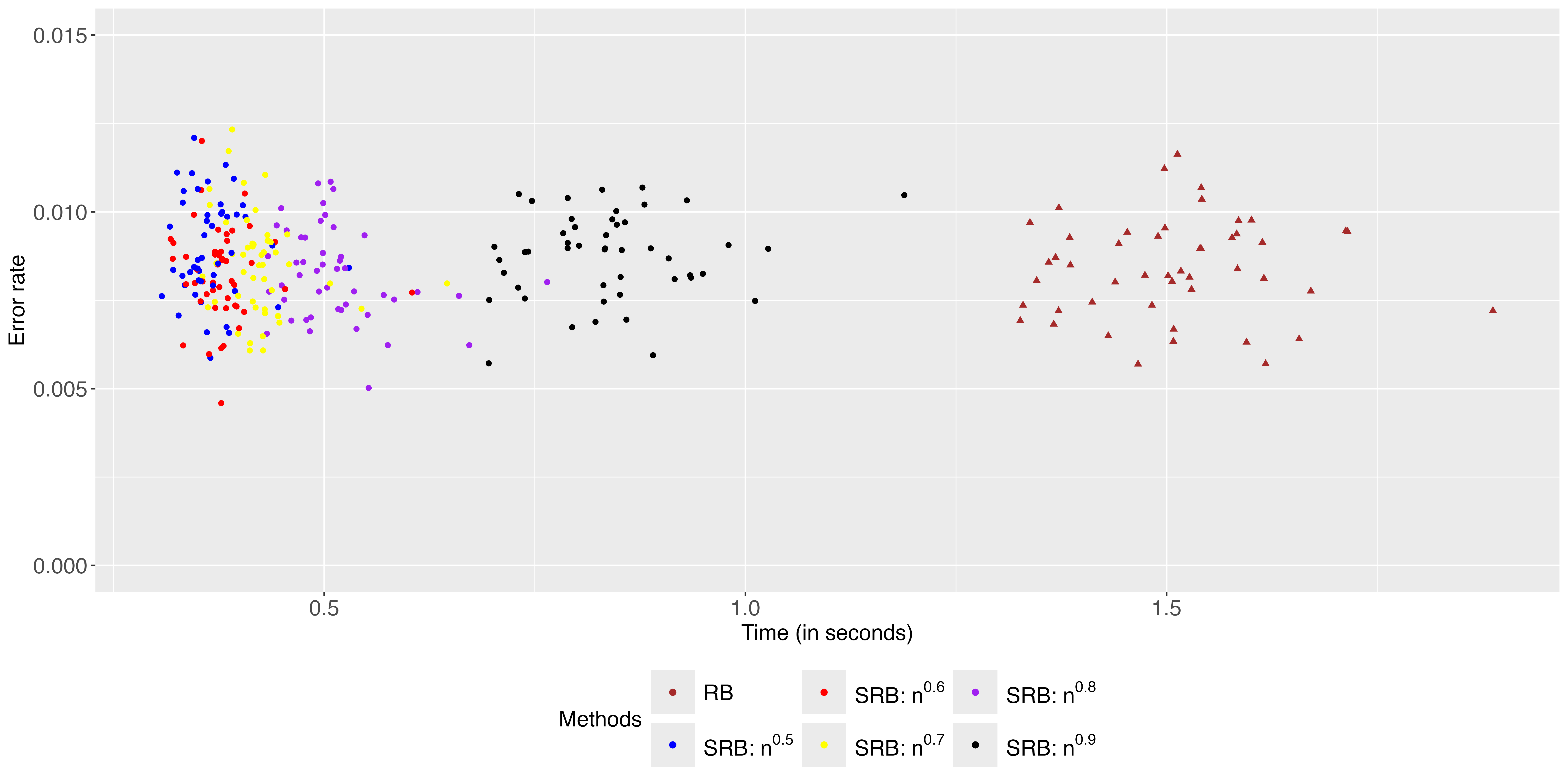}\\
    \includegraphics[height=0.31 \textheight]{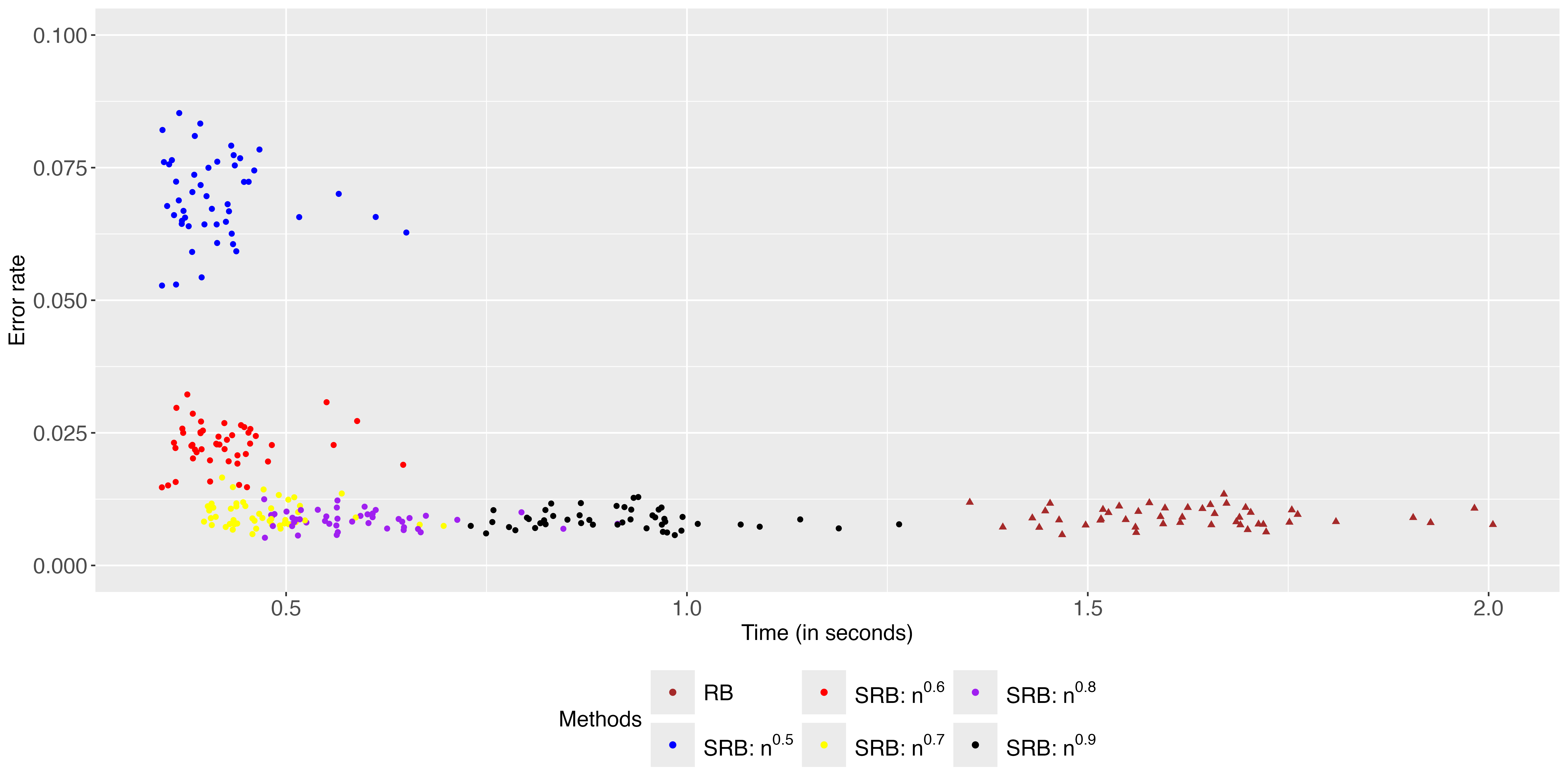}
    \caption{Statistical error vs. runtime (in seconds) for linear model (top), logistic regression (middle), and Poisson regression (bottom).
    The brown triangles denote error rates corresponding to RB, while the other colored dots denote error rates corresponding to SRB, $b=n^{\gamma}$ with $\gamma =0.5,0.6,0.7,0.8,0.9.$ \ig{Here, we have $p=200$ for all three models, $n=50,000$ for logistic and poisson regression and $n=10^5$ for linear model.}
    \label{fig:sim}}
\end{figure}

\begin{table}[h]
\centering
\resizebox{\columnwidth}{!}{%
\begin{tabular}{|c||c|c||c|c||c|c|}
\hline
Model & \multicolumn{2}{|c|}{Linear} &  \multicolumn{2}{|c|}{Logistic} & \multicolumn{2}{|c|}{Poisson} \\ \hline
 &  Error rate in \% & Time in sec & Error rate in \% & Time in sec &  Error rate in \% & Time in sec\\ \hline
RB & 9.66 (1.44) & 4.60  & 0.84 (0.14) & 1.51 & 0.92 (0.17)  & 1.63 \\ \hline
SRB: $n^{0.5}$ & 16.86 (3.95) & 0.71 & 0.90 (0.14)& 0.37 & 6.93 (0.77) & 0.41 \\ \hline
SRB: $n^{0.6}$ & 9.15 (1.45) & 0.69 & 0.82 (0.13) & 0.38  & 2.28 (0.41) & 0.43\\ \hline
SRB: $n^{0.7}$ & 9.29 (1.21) & 0.77 & 0.85 (0.14) & 0.42 & 0.97 (0.24) & 0.47 \\ \hline
SRB: $n^{0.8}$ & 9.46 (1.61) & 0.91 & 0.82 (0.13) & 0.51 & 0.86 (0.16) & 0.58 \\ \hline
SRB: $n^{0.9}$ & 9.64 (1.36) & 1.61 & 0.88 (0.12) & 0.84  & 0.87 (0.17) & 0.92\\ \hline
\end{tabular}%
}
 \caption{\ig{Summary of the results for linear models (left), logistic regression (middle), and Poisson regression (right). For each model, the first column quantifies the performance from a statistical perspective via the mean and standard deviation of error rate.
    Both values are expressed in $\%$, and the standard deviation is reported inside parentheses. The second column for each model provides the average runtime for RB and SRB (for $b=n^{\gamma}$ with $\gamma \in \{0.5,0.6,0.7,0.8,0.9\}$). Here, we have $p=200$ for all three models, $n=50,000$ for logistic and Poisson regression, and $n=10^5$ for linear model.}
    \label{tab:sim_results}}
\end{table}

\begin{figure}[h]
    \centering
    \includegraphics[trim={0cm 3cm 0cm 0cm},clip, scale=0.38]{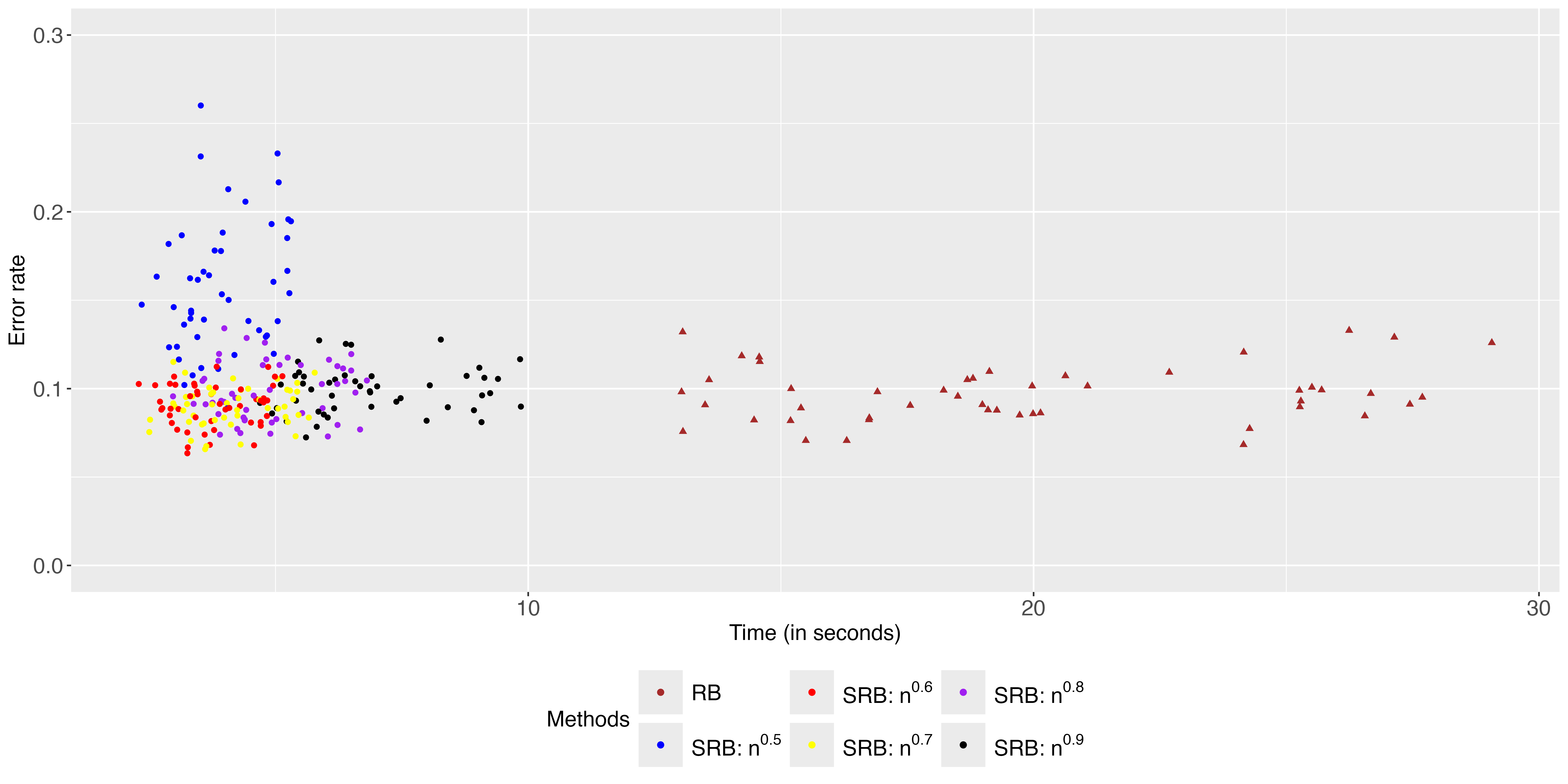}\\
    \includegraphics[trim={0cm 3cm 0cm 0cm},clip, scale=0.38]{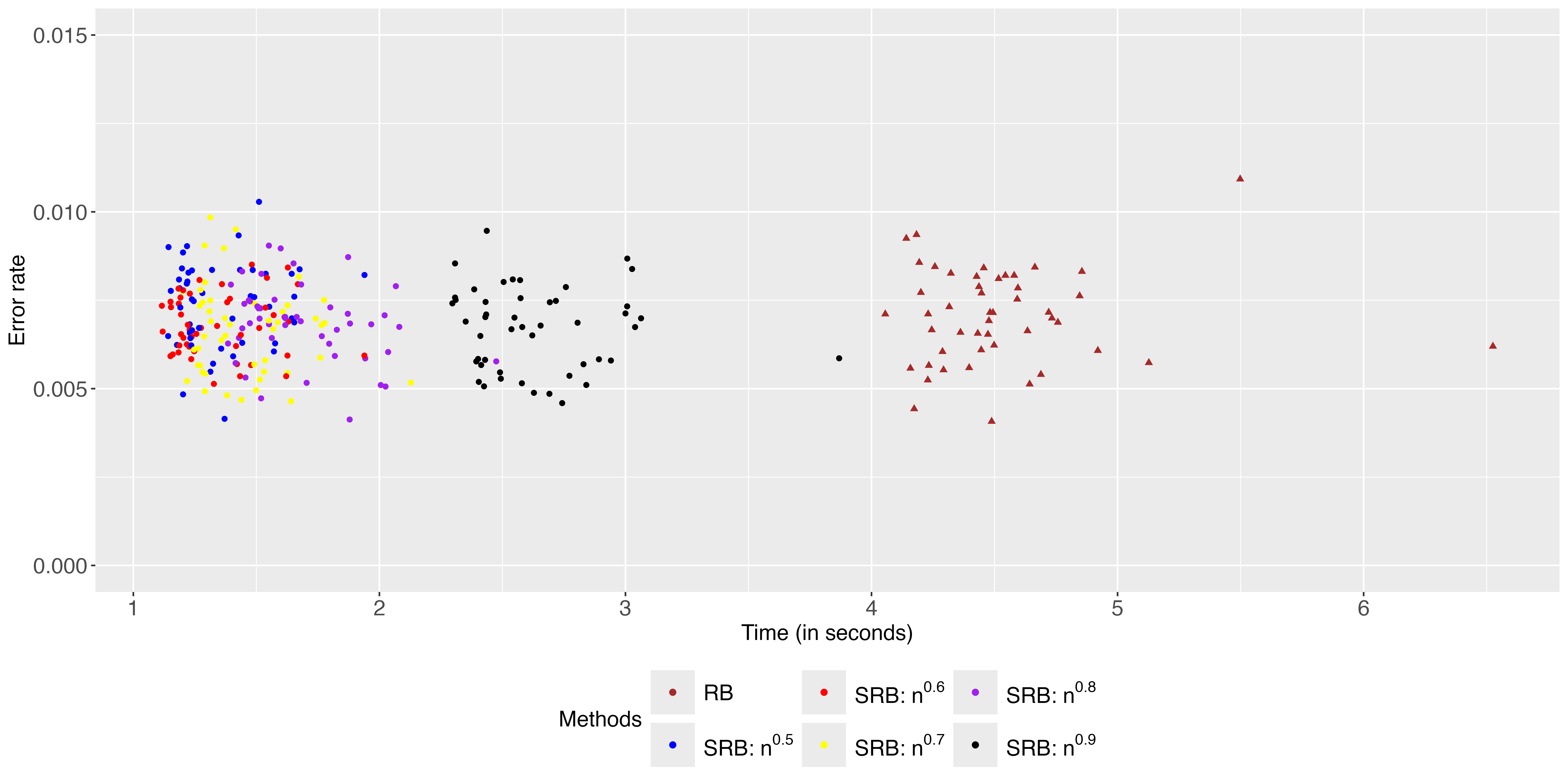}\\
    \includegraphics[height=0.31 \textheight]{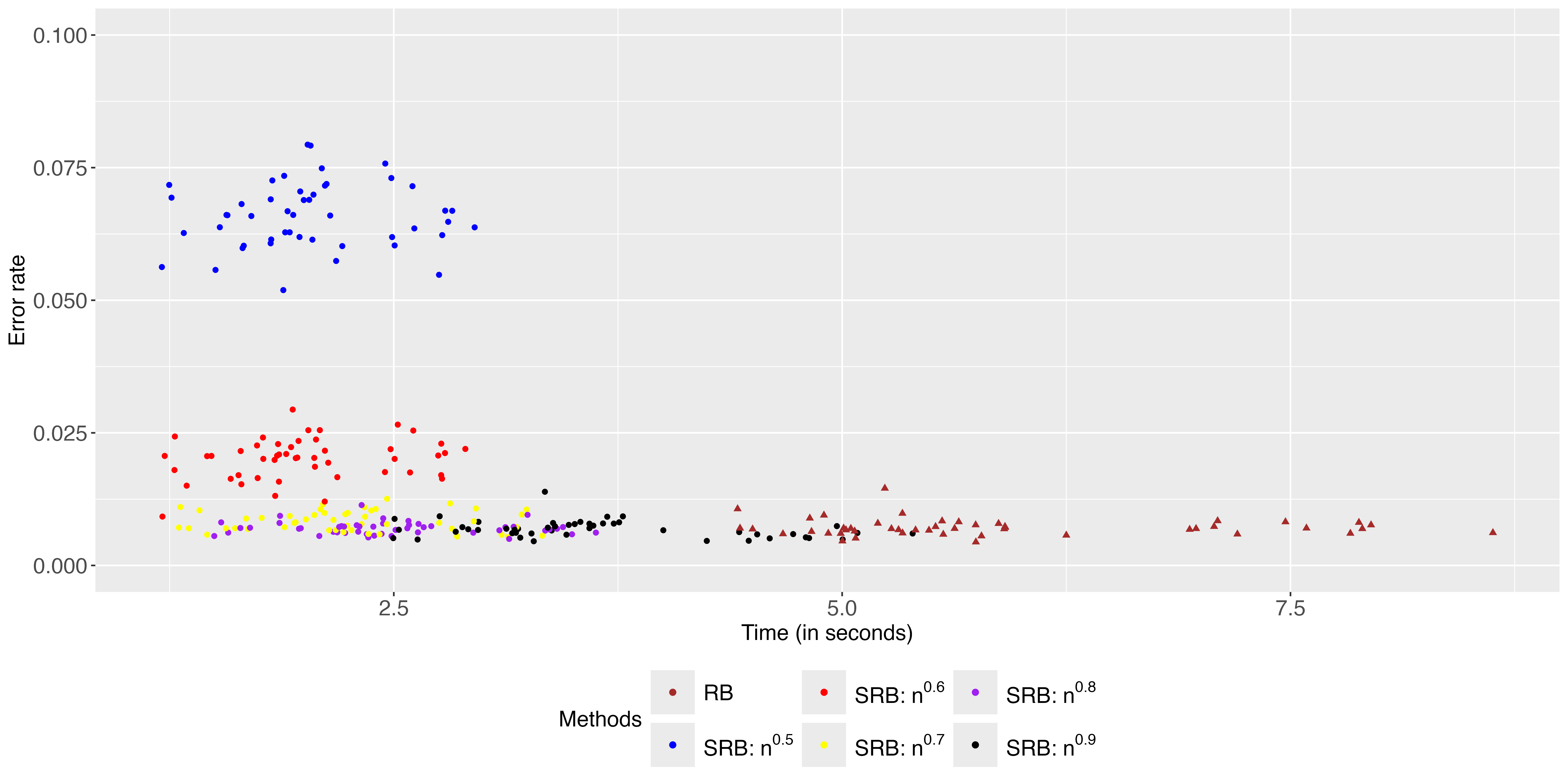}
    \caption{Statistical error vs. runtime (in seconds) for linear model (top), logistic regression (middle), and Poisson regression (bottom).
    The brown triangles denote error rates corresponding to RB, while the other colored dots denote error rates corresponding to SRB, $b=n^{\gamma}$ with $\gamma =0.5,0.6,0.7,0.8,0.9.$ \ig{Here, we have $p=300$ for all three models, $n=10^5$ for logistic and poisson regression and $n=2\times 10^5$ for linear model.}}
    \label{fig:sim_big}
\end{figure}
\begin{table}[h]
\centering
\resizebox{\columnwidth}{!}{%
\begin{tabular}{|c||c|c||c|c||c|c|}
\hline
Model & \multicolumn{2}{|c|}{Linear} &  \multicolumn{2}{|c|}{Logistic} & \multicolumn{2}{|c|}{Poisson} \\ \hline
 &  Error rate in \% & Time in sec & Error rate in \% & Time in sec &  Error rate in \% & Time in sec\\ \hline
RB & 9.72 (1.58) & 19.89  & 0.71 (0.13) & 4.53 & 0.72 (0.16)  & 5.85 \\ \hline
SRB: $n^{0.5}$ & 15.75 (3.74) & 3.95 & 0.74 (0.12)& 1.36 & 6.59 (0.62) & 2.04 \\ \hline
SRB: $n^{0.6}$ & 9.06 (1.22) & 3.77 & 0.68 (0.09) & 1.35  & 2.01 (0.39) & 2.01\\ \hline
SRB: $n^{0.7}$ & 8.97 (1.13) & 4.16 & 0.66 (0.13) & 1.45 & 0.84 (0.19) & 2.27 \\ \hline
SRB: $n^{0.8}$ & 9.84 (1.61) & 4.91 & 0.68 (0.11) & 1.69 & 0.70 (0.12) & 2.50 \\ \hline
SRB: $n^{0.9}$ & 9.91 (1.30) & 6.88 & 0.67 (0.12) & 2.64  & 0.69 (0.16) & 3.63\\ \hline
\end{tabular}%
}
 \caption{\ig{Summary of the results for linear models (left), logistic regression (middle), and Poisson regression (right). For each model, the first column quantifies the performance from a statistical perspective via the mean and standard deviation of error rate.
    Both values are expressed in $\%$, and the standard deviation is reported inside parentheses. The second column for each model provides the average runtime for RB and SRB (for $b=n^{\gamma}$ with $\gamma \in \{0.5,0.6,0.7,0.8,0.9\}$). Here, we have $p=300$ for all three models, $n=10^5$ for logistic and Poisson regression, and $n=2\times 10^5$ for linear model.}
    \label{tab:sim_results_big}}
\end{table}

\subsection{Choice of $b$ in practice}
A relevant question in this context is how to choose the appropriate $b$ in real-world applications. Discussions on the theoretical validity of results established in Sections \ref{subsec:theory_mlr} and \ref{subsec:theory_glm} show that $b>>\sqrt{n}$ is sufficient for the validity of the underlying assumptions, and hence for the theoretical guarantees. 
But how do we choose a $b$ in practice? 

Our recommendation is to choose $b$ as a function of the \textit{relative} time gain that the practitioner is aiming for.
This can be accomplished as follows:
Consider the time gain metric defined as $G = npR/(np+bpR)$, which is approximately the ratio of the total runtimes for RB and SRB.
For a target time gain of $G$ times, $b$ should be set at
$$
b = \frac{npR - Gnp}{bpR}.
$$

To illustrate this idea,
in Fig \ref{fig:speed}, we have plotted the theoretical and the real-world time gain for computing the SRB estimator $\hat{\beta}_{(b)}^*$ for $b=n^{\gamma}$ with $\gamma= 0.65, \ldots, 0.95$, and the RB estimator $\hat{\beta}^*$ for $R=100$ resamples in a linear model setting {with $n=100,000$ and $p=200$}. For the theoretical time gain, we plot the quantity $G = npR/(np+bpR)$ for different choices of $b$. The $X$ axis denotes the choices of $\gamma$ with $\gamma=1$ corresponding to residual bootstrap. The $Y$ axis denotes the ratio of the time taken by RB to the time taken by SRB, both theoretical and real-world. For example, a time gain of 2 denotes that for that particular choice of $\gamma$, SRB is twice as fast as RB. Thus, if we, say, want to improve the computation time of the estimator 10 times, choosing $\gamma$ around $0.8$ seems a reasonable choice with very little compromise in statistical accuracy as can be seen in Figures \ref{fig:sim} and \ref{fig:sim_big} and Tables \ref{tab:sim_results} and \ref{tab:sim_results_big}.
 We observe from the plot that the theoretical time gain metric is a reasonable proxy for the real-world time gain, particularly as $\gamma > 0.65$.
}
\begin{figure}[h]
    \centering
    \includegraphics[height=0.7\textwidth]{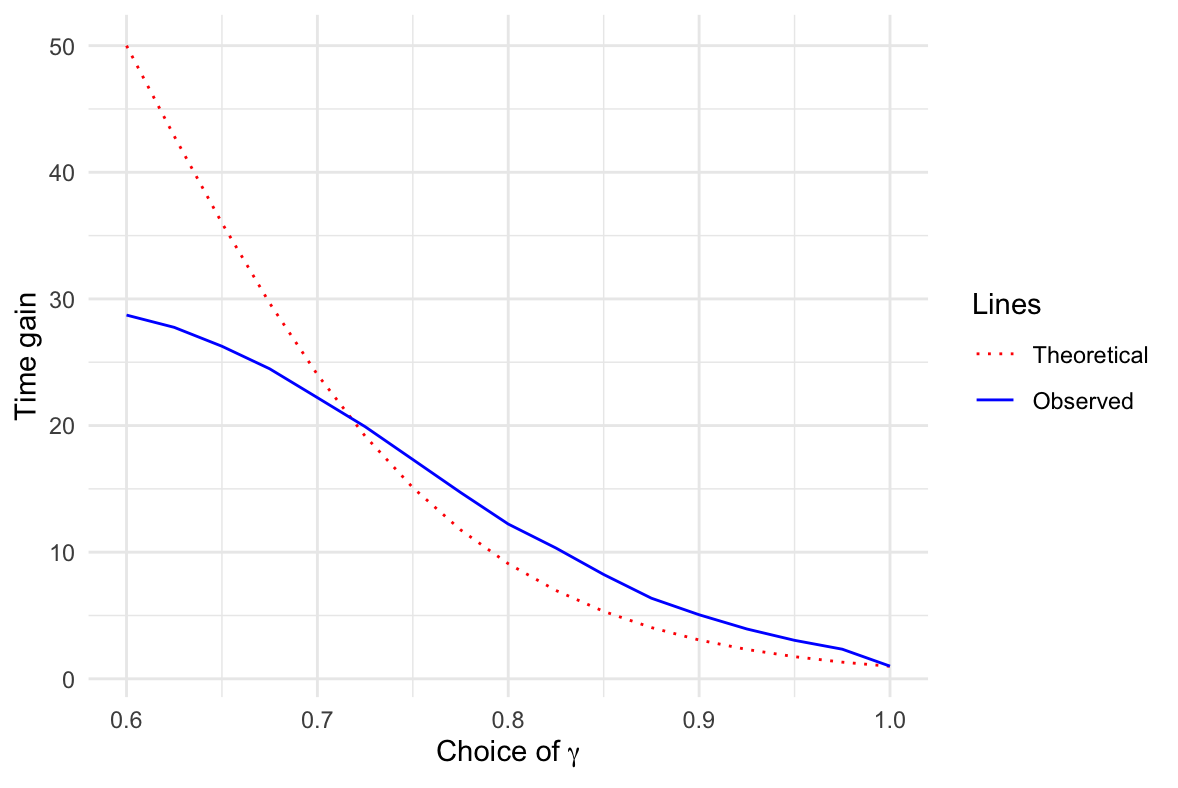}\\
    \caption{Time gain vs $\gamma$ where $b=n^{\gamma}$ with $\gamma =0.6,...,0.95,1$}
    \label{fig:speed}
\end{figure}

\section{Real data analysis}
\label{sec:real}

We used the proposed SRB method to analyze the Forest Cover type data obtained from UCI Machine Learning Repository \citep{misc_covertype_31}.
This data set consists of 581,012 observations corresponding to 7 cover type classes, and 54 other attributes including 10 quantitative variables (Slope, elevation etc.),  4 binary wilderness areas, and 40 binary soil type variables. This data set has been explored in several works, including \cite{blackard1999comparative}, \cite{gama2003accurate}, \cite{oza2001experimental}, and \cite{giannella2005information}. We consider the cover type as the categorical response variable and the remaining 54 variables as covariates. 

We first carried out some pre-processing. Since the categories are heavily unbalanced, we consider a subset of the data involving the two largest cover types, Spruce-Fir and Lodgepole Pine (with a total of $n=495,141$ observations). Once the data set is subsetted, we observe that some of the binary variables are left with a single category (either 0 or 1), and hence, we remove them from the analysis. The modified data set then has 48 variables. Next, we proceed to check if multicollinearity exists among the predictor variables. Although multicollinearity does not affect prediction, it is important to note that the presence of multicollinearity among the predictor variables can result in less precise regression coefficients. One way of handling multicollinearity in regression is to look at the Variance Inflation Factor (VIF) of the predictor variables \citep{midi2010collinearity}. As a rule of thumb, a VIF value greater than 5 indicates high correlation, while values between 1 and 5 denote moderate correlation among the variables. We take a more conservative approach and remove those variables with VIF greater than or equal to 2. For that, we begin by removing the variable with the highest VIF, and recalculating the VIF for all the other predictors. We keep on doing this until all VIF values become less than 2. Thus, we are left with 44 predictor variables, and 4 variables are removed owing to their high VIF values.

Next, we are interested in fitting a multiple logistic regression model of the form $$Y_i \stackrel{ind}{\sim} Ber(p_i) \ \text{where} \ \ logit(p_i)=\beta_0+\sum_{j=1}^p \beta_j X_{ij}$$ to the data,
and estimating $q_{95}$, the $95 \%$ quantile of the root function $\lVert \hat{\beta}-\beta \rVert_2$, which quantifies the precision of the estimator. We applied RB and SRB for this purpose with $100$ resamples. 
For SRB, we used $b= n^{\gamma}$ with $\gamma = 0.55, 0.6, 0.65, \ldots, 0.9$.

\begin{figure}[h]
    \centering
    \includegraphics[trim={1cm 0cm 0cm 0cm},clip,width=0.7\textwidth]{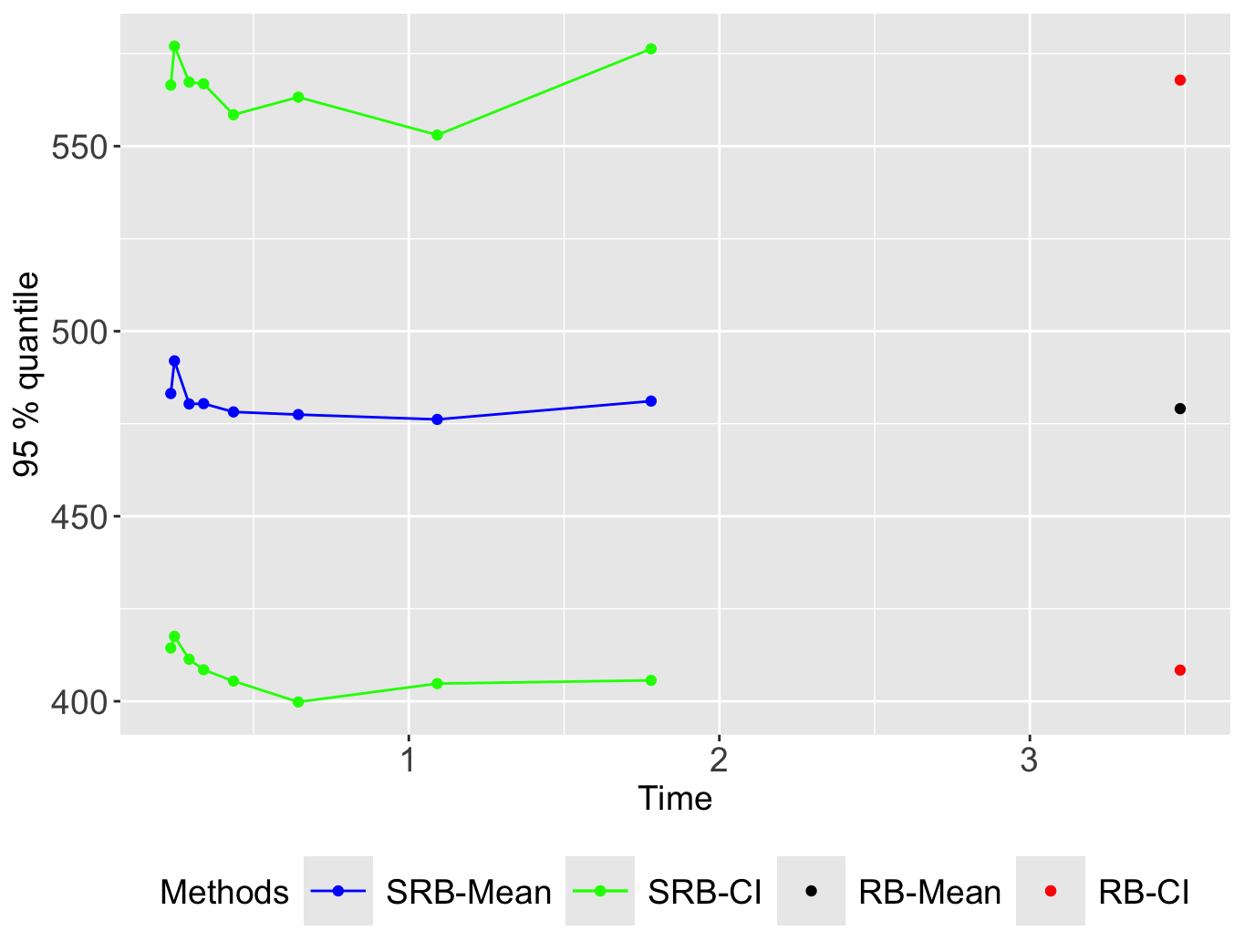}
    \caption{Results from the Covertype data analysis.
    plotted against the time taken. The middle blue line denotes the mean values of $\hat{q}_{95}$ for SRB using $B=500$ replications, while the upper and lower green lines denote the $95 \%$ confidence interval. For SRB, we used $b= n^{\gamma}$ with $\gamma = 0.55, 0.6, 0.65, \ldots, 0.9$.
    The black and red dots denote the mean value and $95 \%$ confidence interval of $\hat{q}_{95}$ for RB.}
    \label{fig:vary_b}
\end{figure}

Since the true value of $q_{95}$ is unknown in this case, we cannot compute the statistical error as we did for the simulation study.
Instead, we compare the estimates from RB and SRB directly.
Figure \ref{fig:vary_b} reports the mean and 95\% confidence interval limits of $\hat{q}_{95}$ obtained from $B=500$ replications of RB and SRB. 
We observe that the estimates from SRB are quite close to those from RB as $\gamma$ increases.
In particular, the average of $\hat{q}_{95}$ from RB (black dot) is well within the 95\% confidence interval from SRB (green curves).
The runtime of SRB is much lower than RB.
This reinforces that SRB can provide a statistical inference of similar quality as RB using significantly lower computational resources.



\section{Discussion} 
\label{sec:disc}
This paper proposes the subsampled residual bootstrap (SRB), which is a faster alternative to classical residual bootstrap. The proposed method preserves the theoretical strengths and practical convenience of residual bootstrap while being computationally more efficient. Moreover, it is highly versatile as it can be applied to any model belonging to the class of generalized linear regression models.
The main idea of our approach is to subsample residuals of size $b=o(n)$ and then form full-size resamples of size $n$ by repeatedly concatenating the subsample. Consistency results for residual bootstrap in GLMs can be obtained as special cases of our derived results for SRB by choosing $b=n$. 

We envision two future directions for research.
First, in this paper, we have focused mainly on the first-order accuracy of SRB. An important next step will be to investigate its higher-order properties.
The results of this investigation will provide us with fine-tuned understanding of the trade-off between statistical accuracy and computational efficiency as a function of $b$

Second, in this paper, we studied the SRB technique under the GLM setting. In future work, it will be interesting to look at how SRB performs in more complex regression models, for example, random forests and decision trees \citep{breiman2001random}.

\clearpage

\appendix
\section{Appendix}
\label{sec:app}


\subsection{Proof of Theorem 1:} Let $\hat{G}_n$ be the empirical distribution which puts mass $\frac{1}{n}$ at $\hat{r}_i$ as defined in Section \ref{sec:mlr} for $i=1, \ldots, n$. Let $E_n()$ denote expectation taken with respect to $\hat{G}_n$. Then, 
    \begin{align*}
        E_n(\hat{\beta}_{(b)}^*)&= E_n( \hat{\beta}+ (X'X)^{-1} X'J' \epsilon_{(b)}^*) \\ &= \hat{\beta}+ (X'X)^{-1} X'J' E_n(\epsilon_{(b)}^*) \\ &= \hat{\beta}+ (X'X)^{-1} X'J' \left( \frac{1}{n} \sum_{i=1}^n \hat{r}_i \right) \\ &= \hat{\beta} \ \ \text{($\hat{r}_i$'s are centered, hence their sum is 0)}
    \end{align*}
    \begin{align*} Var_n(\hat{\beta}_{(b)}^*) &= Var_n( \hat{\beta}+ (X'X)^{-1} X'J' \epsilon_{(b)}^*) \\ &=(X'X)^{-1} X'J'E_n\left( \epsilon_{(b)}^*\right)JX (X'X)^{-1} \\&= \left( \frac{1}{n} \sum_{i=1}^n \hat{r}_i^2 \right) (X'X)^{-1} X'J'JX (X'X)^{-1} \\ &= \frac{\hat{\sigma}_n^2}{n} \left(\frac{X'X}{n}\right)^{-1} \left(\frac{X'J'JX}{n} \right) \left(\frac{X'X}{n}\right)^{-1} \left[\text{where } \hat{\sigma}_n^2= \frac{1}{n} \sum_{i=1}^n \hat{r}_i^2 \right] \\ &\rightarrow 0 \times Q_n^{-1}Q_b Q_n^{-1} =0 
    \end{align*}

Hence, the result follows by a simple application of Chebyshev's inequality.
\subsection{Proof of Theorem 2:}  We first consider the quantity $\sqrt{n}(X'X)^{-1}X'J' \epsilon_{(b)}$ where $\epsilon_{(b)}=(\epsilon_1, \ldots, \epsilon_b)$ is a $b \times 1$ vector with each $\epsilon_i$ having a common distribution $F$. We first state and prove an important result in this context. 
 \begin{result}
 $\sqrt{n}(X'X)^{-1}X'J' \epsilon_{(b)}\xrightarrow{d} N(0, \sigma^2 Q_n^{-1}Q_b Q_n^{-1})$
 \end{result}
 
 \textbf{Proof:} We first note that $\sqrt{n}(X'X)^{-1}X'J' \epsilon_{(b)}=\left(\frac{X'X}{n}\right)^{-1}\frac{1}{\sqrt{n}}X'J'\epsilon_{(b)}$. Consider $p=1$ for simplicity.
 
 Then, let $Z_b=\frac{1}{\sqrt{n}}\sum_{i=1}^b \theta_i \epsilon_i$ where $\theta_i$ is the $i^{th}$ element of $X'J'$ (which is now a row vector).
 Now, $Z_b=\frac{1}{\sqrt{mb}}\sum_{i=1}^b \theta_i \epsilon_i=\frac{1}{\sqrt{m}}\left[\frac{1}{\sqrt{b}} \sum_{i=1}^b \theta_i \epsilon_i  \right]$
 Let $G_i$ be the cdf of $\theta_i \epsilon_i$. Also, let $s_b^2=\sum_{i=1}^b \text{Var} (\theta_i \epsilon_i)= \sigma^2 \sum_{i=1}^b \theta_i^2$. Since this is the scalar case, we have, $$Q_b=\underset{n\rightarrow \infty}{\text{lim}}\frac{1}{n} \sum_{i=1}^b \theta_i^2= \underset{n\rightarrow \infty}{\text{lim}}\frac{1}{mb} \sum_{i=1}^b \theta_i^2$$ [Since $b \rightarrow \infty$ as $n \rightarrow \infty$ by (Assumption \ref{assum_mlr3})]
 First,we try to show that $Z_b= \frac{1}{\sqrt{mb}}\sum_{i=1}^b \theta_i \epsilon_i \xrightarrow{d} N(0,\sigma^2 Q_b)$.
 Hence, according to Lindeberg-Feller Central Limit Theorem, we are first required to show that $$\underset{b\rightarrow \infty}{\text{lim}}\frac{1}{s_b^2}\sum_{i=1}^b \int_{\lvert w\rvert > vs_b} w^2 dG_i(w)=0 \ \ \text{for all } v>0$$
 Now, $G_i(w)=P(\lvert \theta_i \epsilon_i\rvert< w)=P\left(\epsilon_i < \frac{w}{\lvert \theta_i \rvert} \right)= F\left(\frac{w}{\lvert \theta_i \rvert} \right)$. Thus, the above limit becomes $$\underset{b\rightarrow \infty}{\text{lim}}\frac{b}{s_b^2}\sum_{i=1}^b \frac{\theta_i^2}{b}\int_{\lvert w\rvert/\lvert \theta_i \rvert > vs_b/\lvert \theta_i \rvert} \left(\frac{w}{\lvert \theta_i \rvert} \right)^2 dF \left(\frac{w}{\lvert \theta_i \rvert}\right)$$
 Since $$\underset{b\rightarrow \infty}{\text{lim}}\frac{s_b^2}{n}= \sigma^2 \underset{b\rightarrow \infty}{\text{lim}} \frac{1}{n}\sum_{i=1}^b \theta_i^2= \sigma^2 Q_b, \ \ \text{(By Assumption 6)}$$ $$\underset{b\rightarrow \infty}{\text{lim}}\frac{b}{s_b^2}= ( \sigma^2 Q_b)^{-1} \underset{b\rightarrow \infty}{\text{lim}}\frac{b}{n}$$  which is finite. 
 Hence, the only thing which is left to be shown is that $\underset{b \rightarrow \infty}{\text{lim}}\frac{1}{b}\sum_{i=1}^b \theta_i^2 \delta_{i,b}=0$ where $$\delta_{i,b}= \int_{\lvert w\rvert/\lvert \theta_i \rvert > vs_b/\lvert \theta_i \rvert} \left(\frac{w}{\lvert \theta_i \rvert} \right)^2 dF \left(\frac{w}{\lvert \theta_i \rvert}\right)$$
 Now, $\underset{b \rightarrow \infty}{\text{lim}}s_b = \infty \ \ (\text{Since } b\rightarrow \infty)$. Under Assumption \ref{assum_mlr5}, we get $\underset{b \rightarrow \infty}{\text{lim}} \delta_{i,b}=0  \ \ \forall i$. Hence, we prove $$\frac{1}{\sqrt{mb}}\sum_{i=1}^b \theta_i \epsilon_i \xrightarrow{d} N(0,\sigma^2 Q_b)$$ Then, we can conclude that $Z_b \xrightarrow{d} N(0, \sigma^2Q_b)$. Although we proved the result for $p=1$, the result can be proved in a similar way for higher values of $p$. Thus, using Assumption \ref{assum_mlr4} and by the application of Slutsky's Theorem, we prove Result 1.
 
 \vspace{2mm}
 Let $\Psi_n(F)$ be the distribution of $\sqrt{n}(X'X)^{-1}X'J'\epsilon_{(b)}$, that is, a probability distribution in $\mathbb{R}^p$. Let $H$ be an alternative distribution for $\epsilon_i$'s which have mean zero and finite variances. Further, let $d_l^p(\mu,\nu)$ denote the Mallows metric defined as $$d_l^p(\mu,\nu)= \underset{U \in \mu, V \in \nu}{\text{inf}}E^{1/l}\left(\lVert U-V\rVert^l \right)$$ Then we have the following result: \begin{result}
 $d_2^p(\Psi_n(F), \Psi_n(H))^2 \leq n \text{trace} \left[(X'X)^{-1}X'J'JX (X'X)^{-1} \right] d_2 (F,H)^2$
 \end{result}
\textbf{Proof: } Let $A= (X'X)^{-1}X'J'$. Then, $\Psi_n(F)$ is the distribution of $\sqrt{n} A \epsilon_{(b)}$. Similarly, $\Psi_n(H)$ can be defined. Note that $AA'= (X'X)^{-1}X'J'JX(X'X)^{-1}$. Now, we can use Lemma 8.9 of \cite{bickel1981some}. Hence proved.

In Result 2, substituting $\hat{G}_n$ for $H$, we get, $$d_2^p(\Psi_n(F), \Psi_n(\hat{G}_n))^2 \leq n \text{trace} \left[(X'X)^{-1}X'J'JX (X'X)^{-1} \right] d_2 (F,\hat{G}_n)^2$$ Now, $d_2(F, \hat{G}_n) \leq d_2(F, \hat{F}_n)+d_2(\hat{F}_n, \hat{G}_n)$. Using Lemma 2.6 in \cite{freedman1981bootstrapping}, we have $d_2(F, \hat{F}_n) \rightarrow 0$ a.e.

Now, \begin{align*}d_2(\hat{F}_n, \hat{G}_n)&=\underset{\hat{\epsilon}_i}{\text{inf}}\left[E\left\{\left( \hat{\epsilon}_i- \frac{1}{n} \sum_{i=1}^n \hat{\epsilon}_i \right)- \left(\frac{\hat{\epsilon}_i}{\sqrt{1-h_i}}-\frac{1}{n} \sum_{i=1}^n \frac{\hat{\epsilon}_i}{\sqrt{1-h_i}} \right)\right\}^2 \right]^{1/2} \\ &=\underset{\hat{\epsilon}_i}{\text{inf}}\left[E\left\{\hat{\epsilon}_i \left(1-\frac{1}{\sqrt{1-h_i}} \right)-\frac{1}{n} \sum_{i=1}^n \hat{\epsilon}_i \left(1-\frac{1}{\sqrt{1-h_i}} \right)\right\}^2 \right]^{1/2} \\ &\leq \sqrt{2} \underset{\hat{\epsilon}_i}{\text{inf}} \left[E \left(\hat{\epsilon}_i- \frac{\hat{\epsilon}_i}{\sqrt{1-h_i}} \right)^2 + \frac{1}{n^2} E \left(\sum_{i=1}^n\hat{\epsilon}_i-\sum_{i=1}^n \frac{\hat{\epsilon}_i}{\sqrt{1-h_i}} \right)^2\right]^{1/2} \\ & \leq \sqrt{2} \left[d_2 \left(\hat{\epsilon}_i, \frac{\hat{\epsilon}_i}{\sqrt{1-h_i}} \right)+ \frac{1}{n^2} \sum_{i=1}^n d_2 \left(\hat{\epsilon}_i, \frac{\hat{\epsilon}_i}{\sqrt{1-h_i}} \right)  \right] \\  (\text{By Lemma } & 8.6 \text{ of \cite{bickel1981some} and  by using } \sqrt{a^2+b^2} \leq \sqrt{|a|^2}+ \sqrt{|b|^2} \ \text{for } a,b\geq 0 ) \\ &=\sqrt{2} \left[d_2 \left(\hat{\epsilon}_i, \frac{\hat{\epsilon}_i}{\sqrt{1-h_i}} \right)+ \frac{1}{n} d_2 \left(\hat{\epsilon}_i, \frac{\hat{\epsilon}_i}{\sqrt{1-h_i}} \right)  \right] \end{align*}

Now, \begin{align*}d_2 \left(\hat{\epsilon}_i,\frac{\hat{\epsilon}_i}{\sqrt{1-h_i}} \right) & \leq \left[E \left(\hat{\epsilon}_i^2 \left(1-\frac{1}{\sqrt{1-h_i}} \right)^2 \right) \right]^{1/2} \\ &= \left[\frac{(\sqrt{1-h_i}-1)^2}{1-h_i}E(\hat{\epsilon}_i^2) \right]^{1/2} \\ &=\left[\frac{(\sqrt{1-h_i}-1)^2}{1-h_i}(1-h_i)\sigma^2 \right]^{1/2} \\ & \leq \left(\sqrt{1-\frac{1}{n}}-1 \right) \sigma \rightarrow 0 \ \text{as } n \rightarrow \infty.\end{align*}

Thus, $d_2(F, \hat{G}_n)\rightarrow 0$ a.e. Now, $n \text{trace} \left[ (X'X)^{-1} X'J'JX (X'X)^{-1}\right] d_2 (F, \hat{G}_n)^2$ $=$ trace $\left[ \left(\frac{X'X}{n}\right)^{-1} \frac{X'J'JX}{n} \left(\frac{X'X}{n}\right)^{-1}\right] d_2 (F, \hat{G}_n)^2 \rightarrow 0$ a.e. [By Assumption \ref{assum_mlr4}, and due to the fact that $d_2(F, \hat{G}_n)\rightarrow 0$] 

Hence, part (1) is verified. We now introduce, $$\sigma_n^2= \frac{1}{n} \sum_{i=1}^n \epsilon_i^2- \left(\frac{1}{n} \sum_{i=1}^n \epsilon_i \right)^2$$ Clearly, $\sigma_n \rightarrow \sigma$ a.e.

We aim to show first that $\hat{\sigma}_n \rightarrow \sigma$ a.e. Using Lemma 2.4 and 2.7 from \cite{freedman1981bootstrapping}, we have, \begin{align*}
    (\hat{\sigma}_n-\sigma_n)^2 & \leq \frac{1}{n} \sum_{i=1}^n \left(\frac{r_i}{\sqrt{1-h_i}}-\epsilon_i \right)^2 \\& = \frac{1}{n} \sum_{i=1}^n \left(r_i (1-h_i)^{-1/2}-\epsilon_i \right)^2 \\ & \approx \frac{1}{n} \sum_{i=1}^n \left(r_i (1-h_i/2)-\epsilon_i \right)^2 \ \ [\text{Ignoring higher orders of }h_i]  \\& \approx \frac{1}{n} \sum_{i=1}^n (r_i-\epsilon_i)^2 -\frac{1}{n} \sum_{i=1}^n r_i h_i\ \ [\text{Ignoring higher orders of }h_i]  \\& < \frac{1}{n} \sum_{i=1}^n (r_i-\epsilon_i)^2 -\frac{1}{n^2} \sum_{i=1}^n r_i \\ &= \frac{1}{n} \lVert r-\epsilon \rVert^2 -\frac{1}{n^2} \sum_{i=1}^n r_i \\ & \rightarrow 0 \ \  a.e.
\end{align*} Thus, $\hat{\sigma}_n \rightarrow \sigma$ a.e. Hence, part (2) follows by simple application of Slutsky's Theorem.


Before proceeding to (3), we need to prove another result. 

\begin{result}
$E \left(\lVert \hat{\epsilon}^*-J'\epsilon_{(b)}^*\rVert^2 \right)$ is a constant multiple of $\hat{\sigma}_n^2$.
\end{result}
\textbf{Proof:} \begin{align*}
    \hat{\epsilon}^*-J'\epsilon_{(b)}^* &= Y^*-X \hat{\beta}_{(b)}^*-Y^*+X\hat{\beta} \\ &= -X (\hat{\beta}_{(b)}^*-\hat{\beta}) \\ &= -X(X'X)^{-1}X'J'\epsilon_{(b)}^* \\ &= -PJ'\epsilon_{(b)}^*
\end{align*} where $P=X(X'X)^{-1}X'$. Now, \begin{align*}
    E \left(\lVert \hat{\epsilon}^*-J'\epsilon_{(b)}^*\rVert^2 \right) &= E \left((J'\epsilon_{(b)}^*)'P (J'\epsilon_{(b)}^*) \right) \\ &= \text{tr} \left\{P E(J'\epsilon_{(b)}^*)(J'\epsilon_{(b)}^*)' \right\}
\end{align*} [Here,we use the following result: If $E(X)=\mu$ and $Var(X)=\Sigma$, then $E(X'AX)=\mu'A\mu+\text{tr}(A\Sigma)$ where $A$ is a symmetric, idempotent matrix]. Hence, we have, \begin{align*}
   E \left(\lVert \hat{\epsilon}^*-J'\epsilon_{(b)}^*\rVert^2 \right) &= \text{tr}\{P J'E(\epsilon_{(b)}^* \epsilon_{(b)}^{*'}) J \} \\ &= \hat{\sigma}_n^2\text{tr}(PJ'J)
\end{align*} Now, \begin{align*}
    \text{tr}(PJ'J) &= \text{tr}(X(X'X)^{-1}X'J'J) \\ 
                    &= \text{tr} (X'J'JX (X'X)^{-1}) \\ 
                    &= \text{tr} \left(\left(\frac{X'J'JX}{n}\right) \left(\frac{X'X}{n}\right)^{-1}\right) \\
\end{align*} Now, $\left(\frac{X'J'JX}{n}\right) \left(\frac{X'X}{n}\right)^{-1} \rightarrow Q_b Q_n^{-1}$ as $n \rightarrow \infty$ (from Assumption \ref{assum_mlr4}),  which is a fixed matrix, independent of $n$. Hence, $\text{tr}(PJ'J)= k$ (a constant). Hence the result follows.

Let \begin{align*}\sigma_b^{*2}&= \frac{1}{n} \sum_{i=1}^n \epsilon_{i}^{**2}- \left(\frac{1}{n} \sum_{i=1}^n \epsilon_{i}^{**2} \right)^2 \\ &=\frac{1}{b} \sum_{i=1}^b \epsilon_{i}^{*2}- \left(\frac{1}{b} \sum_{i=1}^b \epsilon_{i}^{*2} \right)^2 \end{align*}

Now, \begin{align*}
    E(\lvert \hat{\sigma}_n^{*2}- \sigma_b^{*2}\rvert | Y_1, \ldots, Y_n)^2 & \leq E(( \hat{\sigma}_n^{*2}- \sigma_b^{*2})^2 | Y_1, \ldots, Y_n) \\ & \leq E \left[ \frac{1}{n} \sum_{i=1}^n (\hat{\epsilon}_i^*-\epsilon_i^{**})^2 |Y_1, \ldots, Y_n \right] \ \ (\text{Using Lemma } 2.7 \text{ of \citealt{freedman1981bootstrapping}}) \\ &= E \left[\frac{1}{n} \lVert \hat{\epsilon}^{*2}-J'\epsilon_{(b)}^* \rVert^2 | Y_1,\ldots, Y_n \right] \\ & \leq \hat{\sigma}_n^2 \frac{k}{n} \ \ (\text{Using Result 3}) \\ & \rightarrow 0
\end{align*}
We now need to show that the conditional law of $\sigma_b^{*2}$ is nearly point mass at $\sigma^2$. Conditional on $Y_1, \ldots, Y_n$, we have, \begin{align*}
    d_1 \left(\frac{1}{n}\sum_{i=1}^n \epsilon_i^{**2}, \frac{1}{n}\sum_{i=1}^n \epsilon_i^{2} \right) & \leq \frac{1}{n} E \left(\left\lvert \sum_{i=1}^n (\epsilon_i^{**2}-\epsilon_i^{2}) \right\rvert \right) \\ & \leq \frac{1}{n} \sum_{i=1}^n E \lvert \epsilon_i^{**2}-\epsilon_i^{2} \rvert \\ & = \frac{1}{n} \sum_{i=1}^n d_1 (\epsilon_i^{**2},\epsilon_i^{2} )
\end{align*} Now, $\epsilon_i^{**}$ has conditional distribution $\hat{G}_n$ and $\epsilon_i$ has law $F$ and $d_2 (\hat{G}_n, F) \rightarrow 0$ a.e. (from proof of part 1). 
Hence, $d_1 (\epsilon_i^{**2},\epsilon_i^{2} )\rightarrow 0$ a.e (By lemma 8.5 of \citealt{bickel1981some}) with $\phi(\epsilon)=\epsilon^2$. Thus, the conditional law of $\frac{1}{n} \sum_{i=1}^n \epsilon_i^{**2}$ is nearly equal to the unconditional law of $\frac{1}{n} \sum_{i=1}^n \epsilon_i^{2}$,and hence concentrates near $\sigma^2$. Similarly, the conditional distribution of $\frac{1}{n} \sum_{i=1}^n \epsilon_i^{**}$ concentrates near 0. Hence, (3) is verified.

Part (4) follows directly from (1) and (3) by the application of Slutsky's Theorem.

\subsection{Proof of Theorem 3} Let $E_n()$ denote conditional expectation taken with respect to $\hat{H}_n$. Then,
     \begin{align*}
        E_n(\hat{\beta}_{(b)}^*)&= E_n( \hat{\beta}+ (F_n(\hat{\beta}))^{-1} \Theta' \epsilon_{(b)}^{*}) \\ &= \hat{\beta}+ (F_n(\hat{\beta}))^{-1} \Theta' E_n(\epsilon_{(b)}^{*}) \\ &= \hat{\beta}+ (F_n(\hat{\beta}))^{-1} \Theta'  1 \left( \frac{1}{n} \sum_{i=1}^n \hat{r}_i \right) \\ &= \hat{\beta} \ \ \text{($\hat{r}_i$'s are such that their sum is 0)}
    \end{align*}
    \begin{align*} Var_n(\hat{\beta}_{(b)}^*) &= Var_n( \hat{\beta}+ (F_n(\hat{\beta}))^{-1} \Theta' \epsilon_{(b)}^{*}) \\ &= \left( \frac{1}{n} \sum_{i=1}^n \hat{r}_i^2 \right) (F_n(\hat{\beta}))^{-1} (X'\hat{V}^{1/2}J'J\hat{V}^{1/2}X) (F_n(\hat{\beta}))^{-1} \\ &= \frac{\hat{\sigma}_n^2}{n} \left(\frac{F_n(\hat{\beta})}{n}\right)^{-1} \left(\frac{X'\hat{V}^{1/2}J'J\hat{V}^{1/2}X}{n} \right) \left(\frac{F_n(\hat{\beta})}{n}\right)^{-1} \left[\text{where } \hat{\sigma}_n^2= \frac{1}{n} \sum_{i=1}^n \hat{r}_i^2 \right] \\ &\rightarrow 0 \times M_n^{-1}M_b M_n^{-1} =0 
    \end{align*}

Thus, $\hat{\beta}_{(b)}^*$ is consistent for $\hat{\beta}$.\\
\subsection{Proof of Theorem 4}   We have $\hat{\beta}_{(b)}^*-\hat{\beta}= (X'\hat{V}X)^{-1} \Theta' \epsilon_{(b)}^{*}$. We first consider $p=1$ for simplicity. Then, $\hat{\beta}_{(b)}^*-\hat{\beta}= (X'\hat{V}X)^{-1}\sum_{i=1}^b \theta_i \epsilon_{i}^{*}$ where $\theta_i$ is the $i^{th}$ element of $X'\hat{V}^{1/2}J'$ (which is now a row vector). 

Let $G_i$ be the cdf of $\theta_i \epsilon_{i}^{*}$. Now, let, $$s_b^2=\sum_{i=1}^b \text{Var} (\theta_i \epsilon_i^{*})=\hat{\sigma}_n^2 \sum_{i=1}^b \theta_i^2$$

Since this is the scalar case, $$L=\underset{n \rightarrow \infty}{\lim} \frac{1}{n} \sum_{i=1}^b \theta_i^2=   \underset{b \rightarrow \infty}{\lim} \ \frac{1}{mb} \sum_{i=1}^b \theta_i^2 \ \ [\text{Since as }n \rightarrow \infty, b \rightarrow \infty]$$ 

First we show, $\frac{1}{\hat{\sigma}_n}\left( \sum_{i=1}^b \theta_i^2 \right)^{-1/2} \sum_{i=1}^b \theta_i \epsilon_{i}^{*} \rightarrow N(0,1)$. 

A necessary and sufficient condition for Lindeberg CLT is that $$\underset{b \rightarrow \infty}{\lim} \frac{1}{s_b^2} \sum_{i=1}^b \int_{|\omega|>v s_b} \omega^2 d G_i(\omega)=0 \ \ \forall v>0$$

Now, $G_i(\omega)=P(|\theta_i \epsilon_{i}^{*}|<\omega)=P\left( \epsilon_{i}^{*} < \frac{\omega}{|\theta_i|} \right)= \hat{H}_n \left(\frac{\omega}{|\theta_i|} \right)$.

Then, the limit becomes $$\underset{b \rightarrow \infty}{\lim} \frac{b}{s_b^2} \sum_{i=1}^b \frac{\theta_i^2}{b} \int_{|\omega|/|\theta_i|>v s_b/|\theta_i|} \left( \frac{\omega}{\theta_i}\right)^2 d \hat{H}_n(\frac{\omega}{|\theta_i|})$$

Since $$\underset{b \rightarrow \infty}{\lim} \frac{s_b^2}{n}= \underset{b \rightarrow \infty}{\lim} \left(\frac{\hat{\sigma}_n^2}{n} \sum_{i=1}^b \theta_i^2 \right)= \left( \underset{b \rightarrow \infty}{\lim} \hat{\sigma}_n^2 \right) \left(\underset{b \rightarrow \infty}{\lim} \frac{1}{n}\sum_{i=1}^b \theta_i^2 \right)= M_b \left( \underset{b \rightarrow \infty}{\lim} \hat{\sigma}_n^2 \right) $$ 
we have, $$\underset{b \rightarrow \infty}{\lim}\frac{b}{s_b^2}=\left( M_b\underset{b \rightarrow \infty}{\lim} \hat{\sigma}_n^2 \right)^{-1} \underset{b \rightarrow \infty}{\lim} \frac{b}{n}$$ which is finite as we will show shortly in Result 4 that, unconditionally, $ \hat{\sigma}_n^2 \xrightarrow{p}1$. Hence, we need to show $\underset{b \rightarrow \infty}{\lim} \frac{1}{b} \sum_{i=1}^b \theta_i^2 \delta_{i,b}=0$ where $\delta_{i,b}= \int_{|\omega|/|\theta_i|>v s_b/|\theta_i|} \left( \frac{\omega}{\theta_i}\right)^2 d \hat{H}_n(\frac{\omega}{|\theta_i|})$. Now, $$\underset{b \rightarrow \infty}{\lim} s_b =\infty \text{ as } b \rightarrow \infty.$$

Then, under assumption \ref{assum_glm3}, we have $\underset{b \rightarrow \infty}{\lim} \delta_{i,b}=0 \  \forall i$. Thus, we show, $$\frac{1}{\hat{\sigma}_n}\left( \sum_{i=1}^b \theta_i^2 \right)^{-1/2} \sum_{i=1}^b \theta_i \epsilon_{i}^{*} \rightarrow N(0,1)$$  

In general, for dimension $p$, we have, $$\frac{1}{\hat{\sigma}_n}\left( X'\hat{V}^{1/2}J'J \hat{V}^{1/2} X  \right)^{-1/2} X'\hat{V}^{1/2}J' \epsilon_{(b)}^{*} \xrightarrow{d} N(0,I)$$

Using the fact that, $\hat{\sigma}_n \xrightarrow{p} 1$, we have, $$\left( X'\hat{V}^{1/2}J'J \hat{V}^{1/2} X  \right)^{-1/2} X'\hat{V}^{1/2}J' \epsilon_{(b)}^{*} \xrightarrow{d} N(0,I)$$

From Assumption 6, we have, $\frac{1}{n}X'V^{1/2} J'J V^{1/2} XX \rightarrow L$. Thus, \begin{align*}
   & \left( \frac{X'\hat{V}^{1/2}J'J \hat{V}^{1/2} X}{n}  \right)^{-1/2} \frac{1}{\sqrt{n} }X'\hat{V}^{1/2}J' \epsilon_{(b)}^{*} \xrightarrow{d} N(0,I)\\ & \implies \frac{1}{\sqrt{n} }X'\hat{V}^{1/2}J' \epsilon_{(b)}^{*} \xrightarrow{d} N(0,M_b).
\end{align*}

Now, \begin{align*}
    \sqrt{n}(\hat{\beta}_{(b)}^*-\hat{\beta}) &=\sqrt{n} (X'\hat{V}X)^{-1} X'\hat{V}^{1/2}J'\epsilon_{(b)}^{*} \\ &= \left( \frac{X'\hat{V}X}{n} \right)^{-1} \frac{1}{\sqrt{n}} X'\hat{V}^{1/2}J' \epsilon_{(b)}^{*} \\ & \xrightarrow{d} N(0, M_n^{-1}M_bM_n^{-1})
\end{align*} (Using results 5 and 6  to be stated and proved below).

Hence, we prove part (1). Subsequently, it follows that $$\hat{\beta}_{(b)}^*-\hat{\beta}= (X'\hat{V}X)^{-1}(X'\hat{V}^{1/2}J'J \hat{V}^{1/2}X)^{1/2}(X'\hat{V}^{1/2}J'J \hat{V}^{1/2}X)^{-1/2} X'\hat{V}^{1/2}J' \epsilon_{(b)}^{*}$$ Part (2) follows trivially from this as $\left( X'\hat{V}^{1/2}J'J \hat{V}^{1/2} X  \right)^{-1/2} X'\hat{V}^{1/2}J' \epsilon_{(b)}^{*} \rightarrow N(0,I)$.

Thus, the proof follows. We now prove the three results remaining to be proved.

 \begin{result}
 $\hat{\sigma}_n^2 \xrightarrow{p} 1$
 \end{result} \textbf{Proof:} We have, \begin{align*} \hat{\sigma}_n^2=
     \frac{1}{n}\sum_{i=1}^n \hat{r}_i^2 &= \frac{1}{n} \sum_{i=1}^n \left(\frac{\hat{\epsilon}_i}{\sqrt{\hat{v}_i(1-h_i)}}- \frac{1}{n} \sum_{i=1}^n \frac{\hat{\epsilon}_i}{\sqrt{\hat{v}_i(1-h_i)}} \right)^2 \\ &= \frac{1}{n} \sum_{i=1}^n \left(\frac{y_i-\hat{y}_i}{\sqrt{\hat{v}_i(1-h_i)}}- \frac{1}{n} \sum_{i=1}^n \frac{y_i-\hat{y}_i}{\sqrt{\hat{v}_i(1-h_i)}} \right)^2 \\ &= \frac{1}{n} \sum_{i=1}^n \left(\frac{y_i-b'(\hat{\theta}_i)}{\sqrt{\hat{v}_i(1-h_i)}}\right)^2-\left( \frac{1}{n} \sum_{i=1}^n \left( \frac{y_i-b'(\hat{\theta}_i)}{\sqrt{\hat{v}_i(1-h_i)}} \right) \right)^2
 \end{align*} Now, \begin{align*}
     \frac{1}{n} \sum_{i=1}^n \left(\frac{y_i-b'(\hat{\theta}_i)}{\sqrt{\hat{v}_i(1-h_i)}}\right)^2 = \frac{1}{n} \sum_{i=1}^n \left(\frac{y_i-b'({\theta}_i)}{\sqrt{\hat{v}_i(1-h_i)}}\right)^2 & + \frac{1}{n} \sum_{i=1}^n \left(\frac{b'(\hat{\theta}_i)-b'({\theta}_i)}{\sqrt{\hat{v}_i(1-h_i)}}\right)^2 + \\ &= \frac{2}{n} \sum_{i=1}^n \frac{(y_i-b'({\theta}_i))(b'({\theta}_i)-b'(\hat{\theta}_i))}{(\sqrt{\hat{v}_i(1-h_i)})^2}
 \end{align*} Before proceeding further, we state Kolmogorov's Strong Law of Large Numbers. 
 
\noindent \textbf{Kolmogorov's SLLN:} Assume $X_1, X_2, \ldots$ are independent with means $\mu_1, \mu_2, \ldots$ and variances $\sigma_1^2, \sigma_2^2, \ldots$ such that $\sum_{k=1}^{\infty} \frac{\sigma_k^2}{k^2} < \infty$. Then, $\frac{\sum_{k=1}^n X_k-\sum_{k=1}^n \mu_k}{n} \xrightarrow{\text{a.s.}}0$.
 
 Let us first consider the term $\frac{1}{n} \sum_{i=1}^n \left(\frac{y_i-b'({\theta}_i)}{\sqrt{\hat{v}_i(1-h_i)}}\right)^2$. Now, \begin{align*}
   \sum_{i=1}^n \left(\frac{y_i-b'({\theta}_i)}{\sqrt{\hat{v}_i(1-h_i)}}\right)^2 &=   \sum_{i=1}^n \left(\frac{y_i-b'({\theta}_i)}{\sqrt{{v}_i(1-h_i)}}\right)^2 \left[1+ \left(\frac{v_i}{\hat{v}_i}-1 \right) \right] \\ &=  \sum_{i=1}^n \left(\frac{y_i-b'({\theta}_i)}{\sqrt{{v}_i(1-h_i)}}\right)^2 +  \sum_{i=1}^n \left(\frac{y_i-b'({\theta}_i)}{\sqrt{{v}_i(1-h_i)}}\right)^2 \left( \frac{v_i}{\hat{v}_i}-1 \right)
 \end{align*} We note that $\frac{\left(y_i-b'({\theta}_i) \right)^2}{v_i(1-h_i)}$ for $i=1,2,\ldots$ are independent. Now, \begin{align*}
     E \left[ \frac{\left(y_i-b'({\theta}_i) \right)^2}{v_i(1-h_i)} \right] &= \frac{E(Y_i-b'({\theta}_i))^2}{v_i(1-h_i)}=\frac{v_i}{v_i(1-h_i)}=\frac{1}{1-h_i}.
 \end{align*} \begin{align*}
     \text{Var} \left[\frac{\left(y_i-b'({\theta}_i) \right)^2}{v_i(1-h_i)}\right] &= \frac{1}{v_i^2(1-h_i)^2}E (Y_i-b'({\theta}_i))^4 -  \left[ E \left( \frac{\left(y_i-b'({\theta}_i) \right)^2}{v_i(1-h_i)} \right) \right]^2 \\ &= \frac{b^{(4)}(\theta_i)+3(b''({\theta}_i))^2}{b''({\theta}_i))^2(1-h_i)^2}-\frac{1}{(1-h_i)^2} \\ &= \frac{b^{(4)}(\theta_i)+2(b''({\theta}_i))^2}{b''({\theta}_i))^2(1-h_i)^2} \\ &= \sigma_i^2, \ \ \text{say, where } b^{(n)}() \text{ denotes the } n^{th} \text{ derivative of } b().
 \end{align*}
 Now, $\sum_{i=1}^{\infty} \frac{\sigma_i^2}{i^2}= \sum_{i=1}^{\infty} \frac{b^{(4)}(\theta_i)+2(b''({\theta}_i))^2}{i^2 b''({\theta}_i))^2(1-h_i)^2}= \sum_{i=1}^{\infty}u_i, \ \ \text{say} $.  
 Let $v_i=\frac{1}{i^2}$. Then, $\frac{u_i}{v_i}= \frac{b^{(4)}(\theta_i)+2(b''({\theta}_i))^2}{ b''({\theta}_i))^2(1-h_i)^2}< \infty$ [Since $\frac{1}{n}<h_i<1$ and $b(.)$ is such that all its derivatives exist] Thus, $\sum_{i=1}^{\infty} \frac{\sigma_i^2}{i^2} < \infty$ (By Comparison test of series) 
 Therefore, by Kolmogorov SLLN, we have, $$\frac{1}{n} \sum_{i=1}^n \frac{\left(y_i-b'(\theta_i) \right)^2}{v_i(1-h_i)}- \frac{1}{n} \sum_{i=1}^n \frac{1}{1-h_i} \xrightarrow{a.s.} 0$$
 
 Using the fact that $\hat{\beta}\xrightarrow{p}\beta$, we can say $b'(\hat{\theta}_i)-b'(\theta_i) \xrightarrow{p} 0$ by noting that $b'(\theta_i)=b'(\eta_i)=b'(X_i'\beta)$ is an everywhere continuous function of $\beta$. Hence, we use the continuous mapping theorem. Also, it follows that $\hat{v}_i \xrightarrow{p} v_i$ since $v_i=b''(\theta_i)$ is a continuous function of $\beta$. Thus, we can say   $$\sum_{i=1}^n \left(\frac{y_i-b'(\theta_i)}{\sqrt{{v}_i(1-h_i)}}\right)^2 \left( \frac{v_i}{\hat{v}_i}-1 \right) \xrightarrow{p} 0$$ Also, since $b'(\hat{\theta}_i)-b'(\theta_i) \xrightarrow{p} 0$, we have, $$\frac{1}{n} \sum_{i=1}^n \left(\frac{b'(\hat{\theta}_i)-b'({\theta}_i)}{\sqrt{\hat{v}_i(1-h_i)}}\right)^2 \xrightarrow{p} 0 \text{ and }  \frac{2}{n} \sum_{i=1}^n \frac{(y_i-b'({\theta}_i))(b'({\theta}_i)-b'(\hat{\theta}_i))}{(\sqrt{\hat{v}_i(1-h_i)})^2} \xrightarrow{p} 0$$
 
 Next, we consider the term $\frac{1}{n} \sum_{i=1}^n \left(\frac{y_i-b'(\hat{\theta}_i)}{\sqrt{\hat{v}_i(1-h_i)}}\right)$. Now, \begin{align*}
   \sum_{i=1}^n \left(\frac{y_i-b'(\hat{\theta}_i)}{\sqrt{\hat{v}_i(1-h_i)}}\right) &=   \sum_{i=1}^n \left(\frac{y_i-b'(\hat{\theta}_i)}{\sqrt{{v}_i(1-h_i)}}\right) \left[1+ \left(\frac{v_i}{\hat{v}_i}-1 \right) \right] \\ &=  \sum_{i=1}^n \left(\frac{y_i-b'(\hat{\theta}_i)}{\sqrt{{v}_i(1-h_i)}}\right) +  \sum_{i=1}^n \left(\frac{y_i-b'(\hat{\theta}_i)}{\sqrt{{v}_i(1-h_i)}}\right) \left( \frac{v_i}{\hat{v}_i}-1 \right)
 \end{align*} The term on the right goes to 0 by similar arguments as above. The first term can be rewritten as: $$\sum_{i=1}^n \left(\frac{y_i-b'(\hat{\theta}_i)}{\sqrt{{v}_i(1-h_i)}}\right)= \sum_{i=1}^n \left(\frac{y_i-b'({\theta}_i)}{\sqrt{{v}_i(1-h_i)}}\right)- \sum_{i=1}^n \left(\frac{b'(\hat{\theta}_i)-b'({\theta}_i)}{\sqrt{v_i(1-h_i)}} \right)$$ Again, by similar arguments as above, the last term of the right hand side goes to 0. Now, $$E\left( \frac{Y_i-b'({\theta}_i)}{\sqrt{v_i(1-h_i)}} \right)=0, \ \ \text{Var} \left( \frac{Y_i-b'({\theta}_i)}{\sqrt{v_i(1-h_i)}} \right)= \frac{1}{1-h_i}$$ and $\sum_{i=1}^{\infty} \frac{1}{i^2(1-h_i)} < \infty$ (can be verified by Comparison test). Hence, by Kolmogorov's SLLN, it follows that $\frac{1}{n} \sum_{i=1}^n \left(\frac{y_i-b'({\theta}_i)}{\sqrt{{v}_i(1-h_i)}}\right) \xrightarrow{a.s.} 0$. Again, using the fact that $b'(\hat{\theta}_i) \xrightarrow{p} b'({\theta}_i)$ for each $i$, and by the application of Continuous mapping theorem, it follows that, $$\left(\frac{1}{n} \sum_{i=1}^n \left(\frac{y_i-b'(\hat{\theta}_i)}{\sqrt{\hat{v}_i(1-h_i)}}\right)\right)^2 \xrightarrow{a.s.} 0$$
 
 Thus, we are able to show that $$\frac{1}{n}\sum_{i=1}^n \hat{r}_i^2 -\frac{1}{n} \sum_{i=1}^n \frac{1}{1-h_i} \xrightarrow{p} 0$$ On an average, each $h_i = \bar{h}=\frac{p}{n}$. Under this condition, replacing $h_i$ by $\bar{h}$, we get, as $n \rightarrow \infty$, $\frac{1}{n}\sum_{i=1}^n \hat{r}_i^2 \xrightarrow{p} 1$

\begin{result}
$\frac{F_n(\hat{\beta})}{n}-\frac{F_n(\beta)}{n}\xrightarrow{p} 0$
\end{result} \textbf{Proof:} We have, \begin{align*}
    \frac{F_n(\hat{\beta})}{n}-\frac{F_n(\beta)}{n} &= \frac{1}{n} \sum_{i=1}^n X_iX_i' (\hat{\text{Var}}(Y_i)-\text{Var} (Y_i)) \\ &=\frac{1}{n} \sum_{i=1}^n X_iX_i'( b''(X_i'\hat{\beta})-b''(X_i'{\beta})) 
\end{align*}  The result follows from the fact that $\hat{\beta}\stackrel{\text{a.s.}}{\rightarrow}\beta$, all derivatives of $b(.)$ exist \citep{fahrmeir1985consistency} and $X_i'\beta$ is a continuous function of $\beta$. \begin{result}
$\frac{X'\hat{V}^{1/2}J'J\hat{V}^{1/2}X}{n}-\frac{X'{V}^{1/2}J'J{V}^{1/2}X}{n} \xrightarrow{p} 0$
\end{result} \textbf{Proof:} For simplicity, let us consider $p=1$. Then, after some calculation, the above expression turns out to be: \begin{align*}
    &= \frac{1}{n}\sum_{i=1}^n X_i^2 (\hat{\text{Var}^2(Y_i)}-\text{Var}^2(Y_i)) +  \frac{2}{n} \sum_{i=1}^b \sum_{j=1}^m \sum_{k>j}^m X_{i+3(j-1)}X_{i+3(k-1)} \times\\& \ \ \ \ \ \ \ \ (\hat{\text{Var}(Y_{i+3(j-1)})}\hat{\text{Var}(Y_{i+3(k-1)})}-\text{Var}(Y_{i+3(j-1)})\text{Var}(Y_{i+3(k-1)})) \\ &= \frac{1}{n}\sum_{i=1}^n X_i^2 (b''(X_i'\hat{\beta})^2-b''(X_i'{\beta})^2) +  \frac{2}{n} \sum_{i=1}^b \sum_{j=1}^m \sum_{k>j}^m X_{i+3(j-1)}X_{i+3(k-1)} \times\\& \ \ \ \  (b''(X_{i+3(j-1)}'\hat{\beta})b''(X_{i+3(k-1)}'\hat{\beta})-b''(X_{i+3(j-1)}'{\beta})b''(X_{i+3(k-1)}'{\beta}))
\end{align*}
 The result follows from the fact that $\hat{\beta}\stackrel{\text{a.s.}}{\rightarrow}\beta$, all derivatives of $b(.)$ exist \citep{fahrmeir1985consistency} and $X_i'\beta$ is a continuous function of $\beta$.

 \subsection{Choice of $M$ and $B$ for the simulation study}
 $M$ and $B$ were chosen following the following argument.
Let $\xi_{ij}$ denote the error from the $j^{th}$ iteration of the $i^{th}$ data set for $1\le i \le M$, $1\le j \le B$. and
note that
\begin{align*}
    &SD\left(\frac{1}{MB}\sum_{i=1}^M \sum_{j=1}^B \xi_{ij}\right) = \frac{\sqrt{Var(\xi)}}{\sqrt{MB}}.
\end{align*}
For the $i^{th}$ simulated data set, we can estimate the variance as 
$$s_i^2 =\frac{1}{B-1} \sum_{j=1}^B \left(\xi_{ij}-\bar{\xi_i} \right)^2,$$ 
providing us with the estimated variances $s_1^2$, $s_2^2$, $\ldots, s_M^2$. Thus, we can estimate the variance $Var(\xi)$ using the formula for pooled variance which, in this case, will be given by $$\hat{Var(\xi)}=\frac{1}{M}\sum_{i=1}^M s_i^2 + \frac{1}{M}\sum_{i=1}^M (\bar{\xi_i}-\bar{\xi})^2$$ Let the quantity $SD\left(\frac{1}{MB}\sum_{i=1}^M \sum_{j=1}^B \xi_{ij}\right)$ be `$d$' where `$d$' denotes the acceptable error.  Then, we can get an idea of `$MB$' as $MB= \frac{Var(\xi)}{d^2}$. We can guess $Var(\xi)$ based on a preliminary run of, say, $M_0$ samples, with, say, $B_0$ replications in each. 
Based on this idea, we choose $M=48, B=25$ for each choice of $S$, which ensures that the standard error of the average error rate is below 0.01.

\ig{\subsection{Simulation study: Comparison of SRB vs subsampling for asymptotically pivotal statistic}
In order to illustrate the performance of SRB, it is of interest to compare its performance to other subsampling based methods. As discussed earlier, subsampling based methods explicitly require knowledge of convergence of the estimator and hence our root function of interest as discussed in the previous section is not a suitable candidate for comparison of our method with a Subsampling based method. Hence, in this section, we consider a pivotal quantity as our root function to facilitate such a comparison. Thus, for linear models we consider the root function \begin{equation}
    T_n(\hat{\beta}, \beta)=\lVert (X'X)^{1/2} (\hat{\beta}-\beta)/\sigma \rVert_2
\end{equation} with the rest of the simulation setup exactly as in Section \ref{subsec:sim1}. Noting the results on asymptotic normality established in Sections \ref{subsec:theory_mlr} and \ref{subsec:theory_glm} and the corresponding results  for residual bootstrap and subsampling in \cite{freedman1981bootstrapping}, we estimate the root function by $\lVert(X'X)^{1/2}(\hat{\beta}^*-\hat{\beta})/\hat{\sigma}^* \rVert_2$ for RB, $\lVert(X'J'JX)^{-1/2}(X'X)(\hat{\beta}_{(b)}^*-\hat{\beta})/\hat{\sigma}^* \rVert_2$ for SRB, and $\lVert(X_b'X_b)^{1/2}(\hat{\beta}_{(b)}^*-\hat{\beta})/\hat{\sigma}^*\rVert_2$ for subsampling with $X_b$ denotes the first $b$ rows of $X$.  Note that, although we have used the same notation $\hat{\sigma}_n^*$ to denote the estimated standard deviation based on bootstrap residuals for all three methods, and the notation $\hat{\beta}_{(b)}^*$ to denote the estimated coefficient for SRB and subsampling, they are computed differently depending on the method used. For both logistic and Poisson regression, we consider the root function \begin{equation}
    T_n(\hat{\beta}, \beta)=\lVert (X'VX)^{1/2} (\hat{\beta}-\beta)\rVert_2
\end{equation} which is estimated by $\lVert (X'\hat{V}X)^{1/2}(\hat{\beta}^*-\hat{\beta})\rVert_2$ for RB, $\lVert (X'\hat{V}^{1/2}J'J\hat{V}^{1/2}X)^{-1/2}(X'\hat{V}X)(\hat{\beta}_{(b)}^*-\hat{\beta}) \rVert_2$ for SRB, and $\lVert(X_b'\hat{V}_b X_b)^{1/2}(\hat{\beta}_{(b)}^*-\hat{\beta})\rVert_2$ for subsampling with $\hat{V}_b$ denoting the matrix formed by the first $b$ rows and $b$ columns of $\hat{V}$. Under a similar simulation setup as in Section \ref{subsec:sim1}, we note the average error rates and runtimes in Tables \ref{tab:sim_results_pivot} and \ref{tab:sim_results_pivot_big}.


\begin{table}[h]
\centering
\resizebox{\columnwidth}{!}{%
\begin{tabular}{|c||c|c||c|c||c|c|}
\hline
Model & \multicolumn{2}{|c|}{Linear} &  \multicolumn{2}{|c|}{Logistic} & \multicolumn{2}{|c|}{Poisson} \\ \hline
 &  Error rate in \% & Time in sec & Error rate in \% & Time in sec &  Error rate in \% & Time in sec\\ \hline
RB & 9.66 (1.44) & 4.60  & 0.84 (0.14) & 1.51 & 0.92 (0.17)  & 1.63 \\ \hline
RB & 1.43 (0.22) & 12.66 &  0.89 (0.13) & 2.07  & 0.82 (0.12) & 2.05 \\ \hline
SRB: $n^{0.5}$ & 5.08 (0.32) & 7.23 & 4.28 (0.21)& 0.76  & 1.91 (0.50) & 0.75 \\ \hline
SRB: $n^{0.6}$& 2.12 (0.35) & 7.26 &  1.58 (0.23)& 0.79  & 1.06 (0.29)& 0.77 \\ \hline
SRB: $n^{0.7}$ & 1.58 (0.23) & 7.27 &  1.02 (0.18)& 0.85  & 0.86 (0.13)& 0.85 \\ \hline
SRB: $n^{0.8}$ & 1.46 (0.21)& 7.53 & 0.94 (0.15)& 1.03  & 0.85 (0.14) & 1.02 \\ \hline
SRB: $n^{0.9}$ & 1.47 (0.23)& 8.26 & 0.88 (0.16)& 1.42 & 0.80 (0.13) &  1.41 \\ \hline
Subsamp: $n^{0.5}$ & 82.51 (1.46) & 0.05 & 4.26 (0.20) & 0.03 & 1.89 (0.45) & 0.07 \\ \hline
Subsamp: $n^{0.6}$ & 15.01 (0.76)& 0.1 & 1.60 (0.25) & 0.05 & 1.05 (0.25) & 0.09 \\ \hline
Subsamp: $n^{0.7}$ & 4.35 (0.51) &  0.32 & 1.02 (0.17) & 0.08 & 0.86 (0.15) & 0.16 \\ \hline
Subsamp: $n^{0.8}$ & 1.55 (0.27)& 1.24 & 0.92 (0.15) & 0.24 & 0.82 (0.17) & 0.5\\ \hline
Subsamp: $n^{0.9}$ & 1.34 (0.23)& 4.23 & 0.90 (0.13) & 0.72 & 0.87 (0.15) & 1.5 \\ \hline
\end{tabular}%
}
\caption{\ig{Summary of the results for linear models (left), logistic regression (middle), and Poisson regression (right). For each model, the first column quantifies the performance from a statistical perspective via the mean and standard deviation of error rate.
    Both values are expressed in $\%$, and the standard deviation is reported inside parentheses. The second column for each model provides the average runtime for RB and SRB (for $b=n^{\gamma}$ with $\gamma \in \{0.5,0.6,0.7,0.8,0.9\}$). Here, we have $p=200$ for all three models, $n=50,000$ for logistic and Poisson regression, and $n=10^5$ for linear model.}
    \label{tab:sim_results_pivot}}
\end{table}

\begin{table}[h]
\centering
\resizebox{\columnwidth}{!}{%
\begin{tabular}{|c||c|c||c|c||c|c|}
\hline
Model & \multicolumn{2}{|c|}{Linear} &  \multicolumn{2}{|c|}{Logistic} & \multicolumn{2}{|c|}{Poisson} \\ \hline
 &  Error rate in \% & Time in sec & Error rate in \% & Time in sec &  Error rate in \% & Time in sec\\ \hline
RB & 1.26 (0.21) & 39.45 &  0.72 (0.15) & 5.78  & 0.71 (0.12) & 5.27 \\ \hline
SRB: $n^{0.5}$ & 4.48 (0.25) & 22.66 & 3.78 (0.16)& 2.63  & 1.51 (0.38) & 2.26 \\ \hline
SRB: $n^{0.6}$& 1.76 (0.30) & 22.53 &  1.26 (0.19)& 2.61  & 0.81 (0.18)& 2.19 \\ \hline
SRB: $n^{0.7}$ & 1.28 (0.22) & 22.76 &  0.84 (0.15)& 2.85  & 0.71 (0.13)& 2.38 \\ \hline
SRB: $n^{0.8}$ & 1.22 (0.21)& 23.2 & 0.76 (0.12)& 3.12  & 0.68 (0.13) & 2.65 \\ \hline
SRB: $n^{0.9}$ & 1.24 (0.20)& 23.05 & 0.73 (0.11)& 4.28 & 0.69 (0.11) &  3.81 \\ \hline
Subsamp: $n^{0.5}$ & 89.79 (0.85) & 0.09 & 3.77 (0.17) & 0.06 & 1.54 (0.35) & 0.06 \\ \hline
Subsamp: $n^{0.6}$ & 13.90 (0.56)& 0.26 & 1.25 (0.23) & 0.10 & 0.86 (0.22) & 0.09 \\ \hline
Subsamp: $n^{0.7}$ & 3.65 (0.45) &  1.06 & 0.83 (0.14) & 0.25 & 0.72 (0.12) & 0.21 \\ \hline
Subsamp: $n^{0.8}$ & 1.20 (0.21)& 3.57 & 0.76 (0.12) & 0.63 & 0.71 (0.13) & 0.59\\ \hline
Subsamp: $n^{0.9}$ & 1.12 (0.18)& 11.76 & 0.73 (0.11) & 1.97 & 0.68 (0.13) & 1.87 \\ \hline
\end{tabular}%
}
\caption{\ig{Summary of the results for linear models (left), logistic regression (middle), and Poisson regression (right). For each model, the first column quantifies the performance from a statistical perspective via the mean and standard deviation of error rate.
    Both values are expressed in $\%$, and the standard deviation is reported inside parentheses. The second column for each model provides the average runtime for RB and SRB (for $b=n^{\gamma}$ with $\gamma \in \{0.5,0.6,0.7,0.8,0.9\}$). Here, we have $p=300$ for all three models, $n=10^5$ for logistic and Poisson regression, and $n=2\times 10^5$ for linear model.}
     \label{tab:sim_results_pivot_big}}
\end{table}
}

\clearpage
\noindent

\vskip 0.2in
\bibliography{ref}

\end{document}